\documentclass[preprint,review,12pt]{elsarticle}

\usepackage{graphics}

\usepackage{amssymb}

\usepackage{subeqn}

\usepackage{bm}

\usepackage{multirow}

\usepackage{rotating}

\biboptions{sort&compress}

\journal{Computers $\&$ Mathematics with Applications}

\begin{document}

\begin{frontmatter}



\title{14-velocity and 18-velocity multiple-relaxation-time lattice Boltzmann models for three-dimensional incompressible flows}

\author[a]{Wenhuan Zhang\corref{cor1}}
\cortext[cor1]{Corresponding author, Tel and Fax: +86 574 87608744.}
\ead{zhangwenhuan@nbu.edu.cn}
\author[b]{Baochang Shi}
\author[a]{Yihang Wang}
\address[a]{Department of Mathematics and Ningbo Collabrative Innovation Center of Nonlinear Harzard System of Ocean and Atmosphere, Ningbo University, Ningbo 315211, People's Republic of China}
\address[b]{School of Mathematics and Statistics, Huazhong University of Science and Technology, Wuhan 430074, People's Republic of China}

\begin{abstract}
In this paper, 14-velocity and 18-velocity multiple-relaxation-time (MRT) lattice Boltzmann (LB) models are proposed for three-dimensional incompressible flows. These two models are constructed based on the incompressible LBGK model proposed by He et al. (Chin. Phys., 2004, 13: 40-46) and the MRT LB model proposed by d'Humi\`{e}res et al. (Philos. Trans. R. Soc., A, 2002, 360: 437-451), which have advantages in the computational efficiency and stability, respectively. Through the Chapman-Enskog analysis, the models can recover to three-dimensional incompressible Navier-Stokes equations in the low Mach number limit. To verify the present models, the steady Poiseuille flow, unsteady pulsatile flow and lid-driven cavity flow in three dimensions are simulated. The simulation results agree well with the analytical
solutions or the existing numerical results. Moreover, it is found that the present models show higher accuracy than d'Humi\`{e}res et al. model and better stability than He et al. model.
\end{abstract}

\begin{keyword}
lattice Boltzmann model \sep multiple-relaxation-time \sep three-dimensional incompressible flow
\sep pulsatile flow in 3D rectangular channel

\end{keyword}

\end{frontmatter}


\section{Introduction}
\label{1}The lattice Boltzmann method is an innovative approach based on kinetic theory to simulate
various complex fluid systems \cite{AidunReview,ChenReview}. The significant advantages of lattice Boltzmann method are the natural parallelism
of algorithm, simplicity of programming and ease of dealing with complex boundary conditions. It has been successfully
applied in the field of complex fluids, such as multiphase fluids \cite{MultiphseFlow}, microfluidics
\cite{MicroFluids}, fluids in porous media \cite{PanPorous,FuPorous}, and impinging fluids \cite{ZhangCicp,ZhangCMA}.

Until now, the lattice Bhatnagar-Gross-Krook (LBGK) model, which is based on a single-relaxation-time (SRT)
approximation \cite{BGK1954}, is still the most popular LB model due to its extreme simplicity. The earliest LBGK model is proposed by Qian et al. \cite{QianModel}, which is often used to simulate the incompressible flow in the low Mach number limit. However, through the Chapman-Enskog (C-E) procedure, this model can only recover to the compressible Navier-Stokes (N-S) equations in the low Mach number limit. If the density fluctuation is assumed to be negligible, the incompressible N-S equations can be derived. But in practical simulations, sometimes the density fluctuation cannot be neglected. In this case, there is compressible effect in the simulations and this effect may lead to serious numerical errors. In fact, Qian et al. model can be viewed as a compressible scheme to simulate incompressible flows. There are efforts to weaken the low Mach number restriction of Qian et al. model to extend this model for compressible flows \cite{YanG1,YanG2}, while in our paper we are focused on how to reduce the compressible effect in existing LB models.

In order to reduce the compressible effect in Qian et al. model, many other LB models have been proposed\cite{ZouModel,HeModel,GuoModel,ShiModel}. Among them, the models proposed by He and Luo, and Guo et al. are widely used. The basic idea of He-Luo model \cite{HeModel} is to neglect the terms of higher order Mach number in equilibrium distribution function, which can explicitly reduce the compressible effect as demonstrated in the following simulations. However, He-Luo model can only accurately recover to the incompressible momentum equations, but keeps the term $\frac{1}{c_s^2}\partial p/\partial t$ in the continuity equation, where $p=c_s^2\rho/\rho_0$ is the normalized pressure. When He-Luo model is applied to the unsteady flow, it requires an additional condition $T \gg L/{c_s}$ ($T$ and $L$ are characteristic time and length, respectively), to eliminate the compressible effect.

As we know, the incompressible limit is equivalent to low Mach number limit.
To overcome the shortcoming of He-Luo model, Guo et al. proposed a new LBGK model \cite{GuoModel} for two-dimensional incompressible flows.
Guo et al. model can exactly recover to the incompressible N-S equations only in the low Mach number limit,
which is accomplished by completely decoupling the pressure and density and delicately designing the equilibrium distribution function.
To our knowledge, Guo et al. model is the first LBGK model which can be applied to steady and unsteady incompressible flows while simultaneously eliminating the compressible effect completely. Due to the advantage of Guo et al. model, He et al. extended this model to three dimensions and proposed the three-dimensional 15-velocity and 19-velocity LBGK models for incompressible flows \cite{ShiModel}.

Although the LBGK model has a very simple algorithm and is popularly used, its stability is not always satisfactory in the practical applications.
To overcome this shortcoming, many other LB models have been developed in the past few years
\cite{ELB1,ELB2,TRT1,TRT2,TRT3,HumieresFirstMRT,Lallemand2DMRT,Humieres3DMRT}. Among them, the
MRT LB model \cite{HumieresFirstMRT,Lallemand2DMRT,Humieres3DMRT} has received the most attention due to its superior numerical stability.
d'Humi\`{e}res firstly proposed the MRT LB model \cite{HumieresFirstMRT} at nearly the same time with Qian et al. model.
Lallemand and Luo carried out detailed analysis on this type of model \cite{Lallemand2DMRT} and
found that it has much better performance than the LBGK model in the stability. To further demonstrate the superior stability of MRT model over the LBGK model, d'Humi\`{e}res et al. developed three-dimensional 15-velocity and 19-velocity MRT models \cite{Humieres3DMRT}.

The MRT model has much better stability than the LBGK model, but in the aspect of computational efficiency the MRT model could be about $15\%$ slower than the LBGK counterpart in terms of the number of nodes updated per second \cite{Humieres3DMRT}. The update of one node includes the memory access and the floating-point operations, so the computational efficiency of MRT LB schemes is mostly limited by the memory access quantity and the calculation amount. To improve the computational efficiency of the present MRT models, we propose a new class of three-dimensional MRT models with less memory consumption and calculation amount in this paper.

The MRT model proposed by d'Humi\`{e}res et al. for three-dimensional incompressible flows \cite{Humieres3DMRT} is based on the Qian et al. model or He-Luo model, which both use $q$ discrete velocity directions in $d$ dimensions, i.e., use the D$d$Q$q$ lattice models. However, it is noticed that the incompressible LBGK model proposed by He et al. \cite{ShiModel} only uses $q-1$ discrete velocity directions in the actual computation. Moreover, d'Humi\`{e}res et al. MRT model contains a moment corresponding to kinetic energy square, which does not affect the recovered macroscopic N-S equations. Therefore, based on the He et al. model and d'Humi\`{e}res et al. model, it is possible to construct an MRT model with a $(q-1)\times(q-1)$ transformation matrix, which can reduce the memory consumption and enhance computation efficiency of the existing MRT models in three dimensions.

The above idea is enlightened by the work of Du and Shi, who proposed a two-dimensional 8-velocity incompressible (iD2Q8) MRT model \cite{DuD2Q8MRT} based on Guo et al. LBGK model and the two-dimensional MRT model proposed by Lallemand and Luo. As a continuing work, we propose two three-dimensional MRT models with 14-velocity and 18-velocity based on He et al. LBGK model and d'Humi\`{e}res et al. MRT model in this paper. The general construction method of $(q-1)\times(q-1)$ transformation matrix is presented. Through the C-E procedure, the proposed models can recover to the incompressible N-S equations in the low Mach number limit. The numerical results of unsteady pulsatile flow and cavity flow show that the proposed model is more accurate than d'Humi\`{e}res et al. MRT model and more stable than He et al. LBGK model, where d'Humi\`{e}res et al. MRT model and He et al. LBGK model are two widely used LB models for three-dimensional incompressible flows. As an example, only the 14-velocity model is presented in details in this paper.

The rest of the paper is organized as follows. Section \ref{3DLBGK} briefly describes the three-dimensional 15-velocity incompressible (iD3Q15) LBGK model proposed by He et al.. Section \ref{3DMRT} presents our three-dimensional 14-velocity incompressible (iD3Q14) MRT model. We provides the simulation results for three benchmark problems: the three-dimensional Poiseuille flow, pulsatile flow and cavity flow by using the proposed iD3Q14 MRT model in section \ref{Results_and_discussion}. We also compared some results with those coming from d'Humi\`{e}res et al. D3Q15 MRT model and He et al. iD3Q15 LBGK model. Section \ref{conclusions} concludes this paper. \ref{appendixA} briefly give the derivation of incompressible N-S equations from iD3Q14 MRT model.
\ref{appendixB} outlines the three-dimensional 18-velocity incompressible (iD3Q18) MRT model.


\section{iD3Q15 LBGK model proposed by He et al.}
\label{3DLBGK}
The iD3Q15 LBGK model proposed by He et al. in Ref. \cite{ShiModel} includes 15 discrete velocity directions as follows:
\begin{displaymath}
\{{\bf{c}_0},{\bf{c}_1},\ldots,{\bf{c}_{14}}\}=
\left(
\setlength{\arraycolsep}{0.12cm}
\begin{array}{ccccccccccccccc}
0 & 1 & { - 1} & 0 & 0 & 0 & 0 & 1 & { - 1} & 1 & { - 1} & 1 & { - 1} & 1 & { - 1} \\
0 & 0 & 0 & 1 & { - 1} & 0 & 0 & 1 & 1 & { - 1} & { - 1} & 1 & 1 & { - 1} & { - 1} \\
0 & 0 & 0 & 0 & 0 & 1 & { - 1} & 1 & 1 & 1 & 1 & { - 1} & { - 1} & { - 1} & { - 1} \\
\end{array}
\right)c
\end{displaymath}
where $c=\delta_x/\delta_t$ is the particle velocity and $\delta_x$ and $\delta_t$ are the lattice spacing and time step, respectively.

The evolution equation of the dynamical system is
\begin{equation}\label{ee}
{f_\alpha }({\bm{x}} + {{\bm{c}}_\alpha }{\delta _t},t + {\delta _t}) - {f_\alpha }({\bm{x}},t) =  - \frac{1}{\tau }({f_\alpha }({\bm{x}},t) - f_\alpha ^{(eq)}({\bm{x}},t))
\end{equation}
where $f_{\alpha}(\bm{x},t)$ and $f_\alpha ^{(eq)}(\bm{x},t)$ are the distribution function and equilibrium distribution function of particle with velocity $\bm c_\alpha$ at node $\bm{x}$ and time $t$, $\tau$ is dimensionless relaxation time. For the iD3Q15 LBGK model, the equilibrium distribution function is designed as
\begin{equation}\label{edf}
f_\alpha ^{(eq)} = {\lambda _\alpha } + {\omega _\alpha }({{{\bm{c}_\alpha}\cdot \bm{u}} \over {c_s^2}} +
{{{{({\bm{c}_\alpha}\cdot \bm{u})}^2}} \over {2c_s^4}} - {{|\bm{u}{|^2}} \over {2c_s^2}}),
\end{equation}
where
\begin{equation}\label{lamda_alpha}
{\lambda _\alpha } = \left\{ {\begin{array}{*{20}{l}}
{{\rho _0} + ({\omega _0} - 1)p/c_s^2,\;\;\;  \alpha=0,}\\
{\;\;\;\;\;\;\;{\omega _\alpha }p/c_s^2,\;\;\;\;\;\;\;\;\;\;\;\;  \alpha=1-18,}
\end{array}} \right.
\end{equation}
and
\begin{equation}\label{omega}
{\omega _\alpha } = \left\{ {\begin{array}{*{20}{l}}
{1/3,\;\;\;\;\;\;\alpha  = 0,}\\
{1/18,\;\;\;\;\;\;\alpha  = 1 - 6,}\\
{1/36,\;\;\;\;\alpha  = 7 - 18,}
\end{array}} \right.
\end{equation}
$c_s=c/\sqrt{3}$ is the sound speed, $p$ and $\bm{u}$ are the pressure and velocity, $\rho_0$ is a constant.

The macroscopic flow velocity $\bm u$ and pressure $p$ are obtained by:
\begin{subequations}\label{hgl}
\begin{equation}\label{hgla}
\emph{\textbf{u}} = \sum\limits_{\alpha  = 1}^{14} {\emph{\textbf{c}}_\alpha } f_\alpha, \hfill \\
\end{equation}
\begin{equation}
p = \frac{{c_s^2 }} {{1 - \omega _0 }}[\sum\limits_{\alpha  = 1}^{14} {f_\alpha }  -
\frac{{\omega _0 \left| {\emph{\textbf{u}}} \right|^2 }} {{2c_s^2
}}].
\end{equation}
\end{subequations}
From Eqs. (\ref{hgl}), we can see that the computation of macroscopic quantities only require the distribution functions in 14 discrete velocity directions, so
iD3Q15 LBGK model is actually a 14-velocity incompressible LBGK model.



Through the C-E expansion, the incompressible Navier-Stokes equations can
be recovered as
\begin{subequations}\label{ns}
\begin{equation}
\nabla \cdot{\bm{u}} = 0, \hfill \\
\end{equation}
\begin{equation}
{{\partial {\bm{u}}} \over {\partial t}} + {\bm{u}}\cdot\nabla {\bm{u}} =  - \nabla p + \nu {\nabla ^2}{\bm{u}},
\end{equation}
\end{subequations}
with the kinematic viscosity
\begin{equation}\label{vis_LBGK}
\nu=c_s^2(\tau-1/2)\delta_t,
\end{equation}where $c_s^2=c^2/3$. The above equation connects the fluid property to the parameter of LBGK model.

\section{incompressible D3Q14 MRT model}
\label{3DMRT}
Based on the above iD3Q15 LBGK model and the D3Q15 MRT model proposed by d'Humi\`{e}res et al. \cite{Humieres3DMRT},
we developed a three-dimensional 14-velocity incompressible (iD3Q14) MRT model. This model adopts the discrete velocity directions as
\begin{displaymath}
\{{\bf{c}_1},{\bf{c}_2},\ldots,{\bf{c}_{14}}\}=
\left(
\setlength{\arraycolsep}{0.12cm}
\begin{array}{cccccccccccccc}
1 & { - 1} & 0 & 0 & 0 & 0 & 1 & { - 1} & 1 & { - 1} & 1 & { - 1} & 1 & { - 1} \\
0 & 0 & 1 & { - 1} & 0 & 0 & 1 & 1 & { - 1} & { - 1} & 1 & 1 & { - 1} & { - 1} \\
0 & 0 & 0 & 0 & 1 & { - 1} & 1 & 1 & 1 & 1 & { - 1} & { - 1} & { - 1} & { - 1} \\
\end{array}
\right)c,
\end{displaymath}
which do not contain the discrete velocity direction $\bm c_0$, and we suppose $c=1$ such that the relevant quantities are dimensionless in the following.

The evolution equation of iD3Q14 MRT model is
\begin{equation}\label{mrteq}
  {f_\alpha }(\bm{x} + {\bm{c}_\alpha }{\delta _t},t + {\delta _t}) - {f_\alpha }(\bm{x},t) =  -
  \sum\limits_{i = 1}^{14} {{\Lambda _{\alpha i}}({f_i}(\bm{x},t) - f_i^{(eq)}(\bm{x},t))} {\rm{, }}\;\;\alpha
  {\rm{ = }}1-14,
\end{equation}
where $f_i^{(eq)}(\bm{x},t)$ is defined in Eq. (\ref{edf}) and $\Lambda_{\alpha i}$ is the element of a $14\times14$ collision matrix $\bm \Lambda$.
The above equation can also be written in a vector form:
\begin{equation}\label{mrteqv}
|f(\bm{x} + {\bm{c}_\alpha }{\delta _t},t + {\delta _t})\rangle - |f(\bm{x},t)\rangle =  - \bm \Lambda
(|f(\bm{x},t)\rangle - |{f^{(eq)}}(\bm{x},t))\rangle,
\end{equation}
where $|f(\bm{x},t)\rangle=(f_1(\bm{x},t),f_2(\bm{x},t),\cdots,f_{14}(\bm{x},t))^{'}$ is a column vector, and the
superscript $'$ represents the transpose operator. For the iD3Q14 MRT model, we have defined a $14\times14$ transformation matrix:
{\small
\begin{equation}
\label{T14}
~\mathbf{T}=\left( {\begin{array}{*{20}c}
   1 & 1 & 1 & 1 & 1 & 1 & 1 & 1 & 1 & 1 & 1 & 1 & 1 & 1  \\
   { - 4} & { - 4} & { - 4} & { - 4} & { - 4} & { - 4} & 3 & 3 & 3 & 3 & 3 & 3 & 3 & 3  \\
   1 & { - 1} & 0 & 0 & 0 & 0 & 1 & { - 1} & 1 & { - 1} & 1 & { - 1} & 1 & { - 1}  \\
   { - 4} & 4 & 0 & 0 & 0 & 0 & 1 & { - 1} & 1 & { - 1} & 1 & { - 1} & 1 & { - 1}  \\
   0 & 0 & 1 & { - 1} & 0 & 0 & 1 & 1 & { - 1} & { - 1} & 1 & 1 & { - 1} & { - 1}  \\
   0 & 0 & { - 4} & 4 & 0 & 0 & 1 & 1 & { - 1} & { - 1} & 1 & 1 & { - 1} & { - 1}  \\
   0 & 0 & 0 & 0 & 1 & { - 1} & 1 & 1 & 1 & 1 & { - 1} & { - 1} & { - 1} & { - 1}  \\
   0 & 0 & 0 & 0 & { - 4} & 4 & 1 & 1 & 1 & 1 & { - 1} & { - 1} & { - 1} & { - 1}  \\
   2 & 2 & { - 1} & { - 1} & { - 1} & { - 1} & 0 & 0 & 0 & 0 & 0 & 0 & 0 & 0  \\
   0 & 0 & 1 & 1 & { - 1} & { - 1} & 0 & 0 & 0 & 0 & 0 & 0 & 0 & 0  \\
   0 & 0 & 0 & 0 & 0 & 0 & 1 & { - 1} & { - 1} & 1 & 1 & { - 1} & { - 1} & 1  \\
   0 & 0 & 0 & 0 & 0 & 0 & 1 & 1 & { - 1} & { - 1} & { - 1} & { - 1} & 1 & 1  \\
   0 & 0 & 0 & 0 & 0 & 0 & 1 & { - 1} & 1 & { - 1} & { - 1} & 1 & { - 1} & 1  \\
   0 & 0 & 0 & 0 & 0 & 0 & 1 & { - 1} & { - 1} & 1 & { - 1} & 1 & 1 & { - 1}  \\
 \end{array} } \right),
\end{equation}
}to transform the distribution function into the moment with the linear mapping $\bm m=\mathbf{T}{|f\rangle}$ and $\bm m^{(eq)}=\mathbf{T}{|f^{(eq)}\rangle}$, and simultaneously convert the collision matrix into a diagonal one by $\bm {\hat  \Lambda} = \mathbf{T}\bm \Lambda \mathbf{T}^{-1}$. Thus, we can further write Eq. (\ref{mrteqv}) as
\begin{equation}\label{mrteqvT}
|f({\bm{x}} + {{\bm{c}}_\alpha }{\delta _t},t + {\delta _t})\rangle  - |f({\bm{x}},t)\rangle  =  - {\mathbf{T}^{ - 1}} \bm {\hat  \Lambda}(\bm m(\bm x,t) - {\bm m^{(eq)}}(\bm x,t)),
\end{equation}
where
\begin{equation}\label{moment}
\bm m = (P,e,j_x,q_x,j_y,q_y,j_z,q_z,3p_{xx},p_{\omega \omega},p_{xy},p_{yz},p_{xz},t_{xyz})',
\end{equation}
and the equilibria of the moments are
\begin{subequations}
\label{meq}
\begin{equation}
\left.
\begin{array}{c}
   P^{(eq)}=7p/3+u^2/3, \\
   e^{(eq)}=-7p+u^2,
\end{array}
\right \}
\end{equation}
\begin{equation}
\left.
\begin{array}{c}
  j_x^{(eq)}=u_x, \\
  j_y^{(eq)}=u_y, \\
  j_z^{(eq)}=u_z, \\
\end{array}
\right \}
\end{equation}
\begin{equation}
\left.
\begin{array}{c}
  q_x^{(eq)}=-7u_x/3, \\
  q_y^{(eq)}=-7u_y/3, \\
  q_z^{(eq)}=-7u_z/3, \\
\end{array}
\right \}
\end{equation}
\begin{equation}
\left.
\begin{array}{c}
  3p_{xx}^{(eq)}=3u_x^2-u^2, \\
  p_{\omega \omega}^{(eq)}=u_{y}^2-u_{z}^2, \\
\end{array}
\right \}
\end{equation}
\begin{equation}
\left.
\begin{array}{c}
  p_{xy}^{(eq)}=u_x u_y, \\
  p_{yz}^{(eq)}=u_y u_z, \\
  p_{xz}^{(eq)}=u_x u_z, \\
\end{array}
\right \}
\end{equation}
\begin{equation}
  t_{xyz}^{(eq)}=0.
\end{equation}
\end{subequations}
It should be noted that there are some differences in the definition of moments for iD3Q14 MRT model and d'Humi\`{e}res et al. D3Q15 MRT model.
Firstly, to construct a 14-velocity MRT model, we discard the moment related to kinetic energy square, which does not affect the hydrodynamics significantly, but is defined in d'Humi\`{e}res et al. D3Q15 MRT model. Secondly, based on the iD3Q15 LBGK model, we define a moment $P$, instead of the moment $\rho$, which is used in d'Humi\`{e}res et al. MRT model. From \ref{appendixA}, we will see that $P$ is also a conserved quantity.

The diagonal collision matrix $\bm{\hat \Lambda}$ is
\begin{equation}\label{sD3Q15}
\bm {\hat  \Lambda}\equiv diag(s_1,s_{2}, s_3, s_{4}, s_5, s_{6},s_7, s_{8},
s_{9}, s_{10}, s_{11}, s_{12}, s_{13}, s_{14}),
\end{equation}
where $s_1$, $s_3$, $s_5$ and $s_7$ are the relaxation parameters corresponding to the conserved moments. It should be noted that, the values of these parameters do not affect the recovered macroscopic N-S equations, which are set to be 1.0 in the following simulations. In addition, Eq. (\ref{sD3Q15}) can also be written as
\begin{equation}\label{sD3Q15meaning}
\bm {\hat  \Lambda}\equiv diag(s_c,s_{e}, s_c, s_{q}, s_c, s_{q}, s_c, s_{q},
s_{\nu}, s_{\nu}, s_{\nu}, s_{\nu}, s_{\nu}, s_{t}),
\end{equation}
where $s_c$, $s_e$, $s_q$, $s_\nu$ and $s_t$ are the parameters corresponding to the conserved moment, the moments related to kinetic energy, energy flux, viscous stress tensor and a third-order moment. This form of collision matrix is used in the Appendix A for recovering the incompressible N-S equations. Finally, it should be emphasized that iD3Q14 MRT model can recover to the iD3Q15 LBGK model by setting $\bm{\hat \Lambda}=(1/\tau)\mathbf{I}$, where $\tau$ is the relaxation time of iD3Q15 LBGK model and $\mathbf{I}$ is the unit matrix.

The transformation matrix $\mathbf{T}$ can be obtained by two ways. The first one is similar to that of d'Humi\'{e}res et al. MRT model, which uses the Gram-Schmidt orthogonalization procedure. Since this procedure for D$d$Q$(q-1)$ lattice model in three dimensions has not been given before, we show the details here. The transformation matrix $\mathbf{T}$ is obtained by orthogonalizing the matrix
\begin{equation}\label{transformation_Mbar}
\mathbf{\overline T}  = [c_{\alpha x}^mc_{\alpha y}^nc_{\alpha z}^l]_{14\times14},m,n,l \ge 0,
\end{equation}
where the element is a polynomial of $c_{\alpha x}^mc_{\alpha y}^nc_{\alpha z}^l$, $m,n$ and $l$ are integers. From a physical viewpoint, 14 row vectors of $\mathbf{\overline T}$ correspond to 14 moments of different orders. We have chosen $\mathbf{\overline T}$ as
\begin{equation}\label{Tbar_vectorform}
\mathbf{\overline T}=(|\overline \phi_1\rangle,|\overline \phi_2\rangle,\cdots,|\overline\phi_{14}\rangle)',
\end{equation}
where
\begin{subequations}
\label{m}
\begin{equation}
\left.
\begin{array}{c}
  |\overline \phi_1\rangle_{\alpha}=\left  \| \bm{c}_\alpha \right \|^0, \\
  |\overline \phi_2\rangle_{\alpha}=\left  \| \bm{c}_\alpha \right \|^2,
\end{array}
\right \}
\end{equation}
\begin{equation}
\left.
\begin{array}{c}
  |\overline \phi_3\rangle_{\alpha}=c_{\alpha x}, \\
  |\overline \phi_5\rangle_{\alpha}=c_{\alpha y}, \\
  |\overline \phi_7\rangle_{\alpha}=c_{\alpha z}, \\
\end{array}
\right \}
\end{equation}
\begin{equation}
\left.
\begin{array}{c}
  |\overline \phi_4\rangle_{\alpha}=\left  \| \bm{c}_\alpha \right \|^2c_{\alpha x}, \\
  |\overline \phi_6\rangle_{\alpha}=\left  \| \bm{c}_\alpha \right \|^2c_{\alpha y}, \\
  |\overline \phi_8\rangle_{\alpha}=\left  \| \bm{c}_\alpha \right \|^2c_{\alpha z}, \\
\end{array}
\right \}
\end{equation}
\begin{equation}
\left.
\begin{array}{c}
  |\overline \phi_9\rangle_{\alpha}=3c_{\alpha x}^{2}, \\
  |\overline \phi_{10}\rangle_{\alpha}=c_{\alpha y}^{2}-c_{\alpha z}^{2}, \\
\end{array}
\right \}
\end{equation}
\begin{equation}
\left.
\begin{array}{c}
  |\overline \phi_{11}\rangle_{\alpha}=c_{\alpha x}c_{\alpha y}, \\
  |\overline \phi_{12}\rangle_{\alpha}=c_{\alpha y}c_{\alpha z}, \\
  |\overline \phi_{13}\rangle_{\alpha}=c_{\alpha x}c_{\alpha z}, \\
\end{array}
\right \}
\end{equation}
\begin{equation}
  |\overline \phi_{14}\rangle_{\alpha}=c_{\alpha x}c_{\alpha y}c_{\alpha z},
\end{equation}
\end{subequations}
$\alpha \in \left \{1,2,\cdots,14\right \}$, $\left \| \bm{c}_{\alpha} \right \|=(c_{\alpha x}^2+c_{\alpha
y}^2+c_{\alpha z}^2)^{1/2}$. The corresponding moments for 14 row vectors are the conserved moment ($\overline m_1=\bar P$), the kinetic energy ($\overline m_2=\bar e$), the momentum ($\overline m_{3,5,7}=\overline j_{x,y,z}$), the energy flux ($m_{4,6,8}=\overline q_{x,y,z}$), the viscous stress tensor ($\overline m_9=3\overline p_{xx}$, $\overline m_{10}=\overline p_{\omega\omega}=\overline p_{yy} -\overline p_{zz}, \overline m_{11,12,13}=\overline p_{xy,yz,xz}$) and an third-order moment $\overline m_{14}=\overline t_{xyz}$. These moments can be classified into four types according to the order. Among them, $\overline P$ is a zeroth-order moment, $\overline j_{x,y,z}$ are first-order ones, $\bar e$ and $\overline p_{xx,\omega\omega,xy,yz,xz}$ are second-order ones, $\overline q_{x,y,z}$ and $\overline t_{xyz}$ are third-order ones. However, in the matrix $\mathbf{\overline T}$, 14 row vectors are organized in the order of corresponding tensors, rather than in the order of corresponding moments. The first two rows of $\mathbf{\overline T}$ correspond to $\overline P$ and $\overline e$, which are scalars or zeroth-order tensors. The next six rows correspond to $\overline j_{x}$, $\overline q_{x}$, $\overline j_{y}$, $\overline q_{y}$, $\overline j_{z}$ and $\overline q_{z}$, which are vectors or first-order tensors. The following five rows represent the viscous stress, which is a second-order tensor. The last row represents a third-order tensor.
After the orthogonalization of $\mathbf{\overline T}$, we obtain the transformation matrix $\mathbf{T}$ and the corresponding moments, which have the similar physical meanings.

The above matrix $\mathbf{\overline T}$ can be explicitly written as
{\footnotesize
\begin{equation}\label{Mbar_explicit}
\mathbf{\overline T}  = \left( {\begin{array}{*{20}{c}}
1&1&1&1&1&1&1&1&1&1&1&1&1&1\\
1&1&1&1&1&1&3&3&3&3&3&3&3&3\\
1&{ - 1}&0&0&0&0&1&{ - 1}&1&{ - 1}&1&{ - 1}&1&{ - 1}\\
1&{ - 1}&0&0&0&0&3&{ - 3}&3&{ - 3}&3&{ - 3}&3&{ - 3}\\
0&0&1&{ - 1}&0&0&1&1&{ - 1}&{ - 1}&1&1&{ - 1}&{ - 1}\\
0&0&1&{ - 1}&0&0&3&3&{ - 3}&{ - 3}&3&3&{ - 3}&{ - 3}\\
0&0&0&0&1&{ - 1}&1&1&1&1&{ - 1}&{ - 1}&{ - 1}&{ - 1}\\
0&0&0&0&1&{ - 1}&3&3&3&3&{ - 3}&{ - 3}&{ - 3}&{ - 3}\\
3&3&0&0&0&0&3&3&3&3&3&3&3&3\\
0&0&1&1&{ - 1}&{ - 1}&0&0&0&0&0&0&0&0\\
0&0&0&0&0&0&1&{ - 1}&{ - 1}&1&1&{ - 1}&{ - 1}&1\\
0&0&0&0&0&0&1&1&{ - 1}&{ - 1}&{ - 1}&{ - 1}&1&1\\
0&0&0&0&0&0&1&{ - 1}&1&{ - 1}&{ - 1}&1&{ - 1}&1\\
0&0&0&0&0&0&1&{ - 1}&{ - 1}&1&{ - 1}&1&1&{ - 1}
\end{array}} \right),
\end{equation}}and $\mathbf{T}$ can be obtained by orthogonalizing the 14 row vectors of $\mathbf{\overline T}$ in an order. Supposing $\textbf{T}=(|\phi_1\rangle,|\phi_2\rangle,\cdots,|\phi_{14}\rangle)'$, then we have
\begin{subequations}
\label{m}
\begin{equation}
\left.
\begin{array}{c}
  |\phi_1\rangle_{\alpha}=\left  \| \bm{c}_\alpha \right \|^0, \\
  |\phi_2\rangle_{\alpha}=7\left  \| \bm{c}_\alpha \right \|^2/2-15\left  \| \bm{c}_\alpha
  \right \|^0/2,
\end{array}
\right \}
\end{equation}
\begin{equation}
\left.
\begin{array}{c}
  |\phi_3\rangle_{\alpha}=c_{\alpha x}, \\
  |\phi_5\rangle_{\alpha}=c_{\alpha y}, \\
  |\phi_7\rangle_{\alpha}=c_{\alpha z}, \\
\end{array}
\right \}
\end{equation}
\begin{equation}
\left.
\begin{array}{c}
  |\phi_4\rangle_{\alpha}=(5\left  \| \bm{c}_\alpha \right \|^2-13)c_{\alpha x}/2, \\
  |\phi_6\rangle_{\alpha}=(5\left  \| \bm{c}_\alpha \right \|^2-13)c_{\alpha y}/2, \\
  |\phi_8\rangle_{\alpha}=(5\left  \| \bm{c}_\alpha \right \|^2-13)c_{\alpha z}/2, \\
\end{array}
\right \}
\end{equation}
\begin{equation}
\left.
\begin{array}{c}
  |\phi_9\rangle_{\alpha}=3c_{\alpha x}^{2}-\left  \| \bm{c}_\alpha \right \|^2, \\
  |\phi_{10}\rangle_{\alpha}=c_{\alpha y}^{2}-c_{\alpha z}^{2}, \\
\end{array}
\right \}
\end{equation}
\begin{equation}
\left.
\begin{array}{c}
  |\phi_{11}\rangle_{\alpha}=c_{\alpha x}c_{\alpha y}, \\
  |\phi_{12}\rangle_{\alpha}=c_{\alpha y}c_{\alpha z}, \\
  |\phi_{13}\rangle_{\alpha}=c_{\alpha x}c_{\alpha z}, \\
\end{array}
\right \}
\end{equation}
\begin{equation}
  |\phi_{14}\rangle_{\alpha}=c_{\alpha x}c_{\alpha y}c_{\alpha z}.
\end{equation}
\end{subequations}The first four orthogonal vectors are related to $P$ and $x$-, $y$- and $z$-momentum modes: $|\phi_1\rangle_{\alpha}=|\overline \phi_1\rangle_{\alpha}$, $|\phi_3\rangle_{\alpha}=|\overline \phi_3\rangle_{\alpha}$, $|\phi_5\rangle_{\alpha}=|\overline \phi_5\rangle_{\alpha}$ and $|\phi_7\rangle_{\alpha}=|\overline \phi_7\rangle_{\alpha}$. The vector $  |\phi_2\rangle_{\alpha}$ is constructed by orthogonalizing the energy mode ${\left| {{\overline \phi _2}} \right\rangle _\alpha }$. Similarly, vectors $|\phi_4\rangle_{\alpha}$, $|\phi_6\rangle_{\alpha}$ and $|\phi_8\rangle_{\alpha}$ are respectively derived upon ${\left| {{\overline\phi _4}} \right\rangle _\alpha }$, ${\left| {{\overline\phi _6}} \right\rangle _\alpha }$ and ${\left| {{\overline\phi _8}} \right\rangle _\alpha }$. $|\phi_9\rangle_{\alpha}$ is built upon $|\overline \phi_9\rangle_{\alpha}$ and $|\phi_{10-14}\rangle_{\alpha}=|\overline \phi_{10-14}\rangle_{\alpha}$. It should be noted that, the row vectors in $\mathbf{T}$ are mutually orthogonal, but they are not normalized to assure that the elements of row vectors are integral, which can simplify the computation. Finally, from above derivation, we can see that the orthogonalization procedure for D$d$Q$(q-1)$ lattice model is the same with that of D$d$Q$q$ lattice model, which promotes the generation of the second way to obtain the transformation matrix $\mathbf{T}$.

The second way to obtain the transformation matrix $\mathbf{T}$ is more straightforward. This way properly uses the orthogonalization procedure, which has been done by d'Humi\`{e}res et al.. The steps are as follows. First, we discard the first column and the third row of the transformation matrix for d'Humi\`{e}res et al. D3Q15 MRT model, which means the distribution function $f_0(\bm x, t)$ in discrete velocity direction $\bm c_0$ and the moment $\epsilon$ related to the kinetic energy square are not used. Secondly, to make the new 14 row vectors orthogonal to each other and assure the elements are integral, we can obtain the transformation matrix $\mathbf{T}$ in our model.

ID3Q14 MRT model is described in details above. Then through the C-E expansion, we can prove that the above model can recover to the incompressible N-S equations as Eq. (\ref{ns}) in the low Mach number limit with the kinematic viscosity
\begin{equation}\label{vis}
\nu=c_s^2(\tau-1/2)\delta_t,
\end{equation}
where $\tau=1/s_{\nu}$, $c_s^2=1/3$ (see Appendix A for details).


\section{Numerical results and discussion}
\label{Results_and_discussion}
In this section, to verify the accuracy and stability of the proposed iD3Q14 MRT model, the three-dimensional (3D) Poiseuille flow and pulsatile flow in a square channel and the 3D lid-driven cavity flow are simulated. In the simulations, the relaxation parameters in the collision matrix $\hat \Lambda$ are chosen to be $s_{c}=1.0$, $s_e=1.19$, $s_{q}=1.2$, $s_{t}=0.98$, $s_{\nu}=1/\tau$, where $\tau$ is related to the viscosity by Eq. (\ref{vis}). For the boundary condition implementation, the non-equilibrium extrapolation scheme is always applied
\cite{GuoBoundary1}. In addition, $c$ is not always set to be 1 in the simulations. In this situation, $u$ and $p$ are firstly normalized with $c$ and $c^2$ and then used to calculate. After the computation is finished, $u$ and $p$ are multiplied by $c$ and $c^2$ as the final results. The relationship between the viscosity and relaxation time is the same with Eq. (\ref{vis}), but $c_s^2=c^2/3$.

\subsection{Steady flow: 3D Poiseuille flow}
\label{4.1}The illustration of three-dimensional Poiseuille flow in a square channel is shown in Fig. \ref{poiseuille-schematic}. The origin $O$ is located at the center of the entrance
plane. The flow region is defined in a rectangular region: $0\leq x \leq l$, $-a\leq y \leq a$ and $-b\leq z \leq b$, where $l=2, a=b=0.5$. Given the boundary condition:
\begin{subequations}\label{bc}
\begin{equation}
\textbf{u}(x,\pm a,z,t) = \textbf{u}(x,y,\pm b,t)=0,
\end{equation}
\begin{equation}
p(0,y,z,t) = {p_{in}},\;\;\;{\mkern 1mu} p(l,y,z,t) = {p_{out}},
\end{equation}
\end{subequations}
where $p_{in}$ and $p_{out}$ are the pressure at the entrance and exit of the channel, the three-dimensional Poiseuille flow has a steady solution \cite{3DPoseiulleFlow},
\begin{subequations}
\begin{equation}
u(y,z,t) = {{16{a^2}} \over {\nu {\pi ^3}}}( - {{dp} \over {dx}})\sum\limits_{i = 1,3,5,...}^\infty  {{{( - 1)}^{(i - 1)/2}}} [1 - {{cosh(i\pi z/2a)} \over
{cosh(i\pi b/2a)}}] \times {{cos(i\pi y/2a)} \over {{i^3}}},
\end{equation}
\begin{equation}
v(x,y,z,t) = w(x,y,z,t) = 0,
\end{equation}
\begin{equation}
   {\mkern 1mu} p(x,t) = {p_{in}} + {{dp} \over {dx}}x,
\end{equation}
\end{subequations}
where $dp/dx=(p_{out}-p_{in})/l$ is the pressure gradient in the channel, $\nu$ is the kinematic viscosity of fluid.

In the simulations, the initial and boundary conditions are set as
\begin{equation}\label{ic}
  \textbf{u}(x,y,z,0) = 0, \;\;\;{\mkern 1mu} p(0<x<l,y,z,0) = {{p_{in}+p_{out}}\over {2}},
\end{equation}
where $p_{in}=1.1$ and $p_{out}=1.0$. To get the numerical solution, the criterion of steady state as follows is used,
\begin{equation}\label{sc}
 {{\sum\limits_{i} | u(\bm x_i,t + \delta t) - u(\bm x_i,t)|} \over {\sum\limits_{i} | u(\bm x_i,t)|}} \le 1.0 \times {10^{ - 10}},
\end{equation}
where $\sum\limits_{i} {} $ denotes the summation over the entire system.

The simulations were carried out with the grid of $65\times33\times33$ for $\nu=0.03$. Fig. \ref{plotmu} shows the variation of $u$ with $y$ for different $z$ location at section $x=1$ and the variation of $p$ with $x$ for different $y$ location at section $z=0$. To demonstrate the numerical results are independent of $\tau$, different $\tau$ ($1/\tau=0.8,1.0,1.2$) are used to simulate the flow. It can be seen from Fig. \ref{plotmu} that, the numerical results are in excellent agreement with the analytical solutions for different $\tau$.


In Fig. \ref{plotmu}, the velocity and pressure are taken from $x=1$ and $z=0$ sections, respectively. To validate the agreement is independent of sections, the variation of $u$ with $y$ at different $x$ sections and the variation of $p$ with $x$ at different $z$ sections are shown in Fig. \ref{plotp}. It can be seen that, the
numerical results are still in excellent agreement with the analytical solutions.

We also carried out the simulations with different lattice spacings. The global relative error in velocity field ($GRE_u$) between the numerical result and the analytical solution is defined as
\begin{equation}\label{ed}
  GRE_{u} = {\sqrt{\sum\limits_i[{(u_{n}-u_{a})^2+(v_{n}-v_{a})^2+(w_{n}-w_{a})^2}]}\over {\sqrt{\sum\limits_i[{u_{a}^2+v_{a}^2+w_{a}^2}]}}},
\end{equation}
where $u_{n}$, $v_{n}$, $w_{n}$ and $u_{a}$, $v_{a}$, $w_{a}$ denote the numerical and analytical velocities, respectively, and the summation of $i$ is over every grid point. The $GRE_u$ for different lattice spacings at different $1/\tau$ are displayed in Table \ref{GREu}. The variation of $GRE_u$ with the lattice spacing is
also shown in Fig. \ref{edx}. The slopes of fitting lines are about 1.85, 2.00 and 1.80 for $1/\tau=$0.8, 1.0 and 1.3, respectively. This shows that the proposed iD3Q14 MRT model is of second order accuracy in space when simulating steady 3D Poiseuille flow.

\subsection{Unsteady flow: 3D pulsatile flow}
\label{4.2}The 3D pulsatile flow is used to validate the proposed model for unsteady flow. The geometric configuration of 3D pulsatile flow is the same with that of 3D Poiseuille flow, but the flow is driven by a periodic pressure gradient between the two ends of the channel.

It should be noted that the 3D pulsatile flow in a circular pipe \cite{3DPulsatileFlow_GuoZL,3DPulsatileFlow_ZhangTing} is usually simulated by the lattice Boltzmann method, while the 3D pulsatile flow in a rectangular channel is little done before. We found an analytic solution for the 3D pulsatile flow in a rectangular channel in a very early paper \cite{3DPulsatileFlow}. We hope this flow and its analytic solution can be widely used for validating the 3D LB models in the future.



Supposing that the flow in the rectangular channel is laminar and incompressible, then the incompressible N-S equations for this flow are reduced to
\begin{equation}\label{pulsatile-ns}
\frac{{\partial u}}{{\partial t}} =  - \frac{{\partial p}}{{\partial x}} + \nu (\frac{{{\partial ^2}u}}{{\partial {y^2}}} + \frac{{{\partial ^2}u}}{{\partial {z^2}}}),
\end{equation}
with the following pressure gradient imposed on the flow
\begin{equation}\label{pbc-pul}
{{dp}\over{dx}}=Gcos(\omega t),
\end{equation}
where $G$ is an amplitude and $\omega$ is a frequency. The analytical solution of above flow is \cite{3DPulsatileFlow}:
\begin{equation}\label{wa}
{u}(y,z,t) = {\rm{Re}}\left\{ {i\frac{G}{\omega }\left\{ \begin{array}{l}
1 - 2\sum\limits_{n = 0}^\infty  {\frac{{{{( - 1)}^n}}}{{{p_n}}}} [\frac{{\cosh ({\gamma _n}y/b)\cos ({p_n}z/b)}}{{\cosh ({\gamma _n}a/b)}} + \\
\frac{{\cosh ({\sigma _n}z/b)\cos ({q_n}y/b)}}{{\cosh ({\sigma _n})}}]
\end{array} \right\}{e^{i\omega t}}} \right\},
\end{equation}
where
\begin{subequations}
\begin{equation}
{\gamma _n} = \sqrt {p_n^2 + i{\eta ^2}} ,\;\;\;\;{\sigma _n} = \sqrt {q_n^2 + i{\eta ^2}},
\end{equation}
\begin{equation}
{p_n} = {{2n + 1} \over 2}\pi ,\;\;\;\;{\rm{  }}{q_n} = {{2n + 1} \over 2}\pi {b \over a},
\end{equation}
\end{subequations}
$\eta  = b\sqrt {\omega /\nu }$ is the Womersley number. In the present study, the flow region is defined in a square channel with $a=b=0.5$.

The simulation parameters are set as follows. The grid size is $Nx\times Ny\times Nz= 81\times 41\times 41$, the period of the changing pressure is $T= 100$ ($\omega = 2\pi/T$), and the magnitude of total pressure drop along the channel is $\Delta p = 0.001$ ($G =\Delta p/l, l=2$), and the pressure at the
outlet ($p_{out}$) is set to be $1.0$. $\delta t=0.0125$ is fixed in the simulations so that one period contains integral step. The initial state of velocity field is always set to be zero everywhere while the initial state of the pressure field is always set to be $(p_{in}+p_{out})/2$, where $p_{in}=p_{out}-\Delta p$ is the pressure at the entrance ($p_{in}$ is determined by the Eq. (\ref{pbc-pul}) at
$t=0$). The calculation of velocity field always began with several periods of initial steps to attain convergence criterion:
\begin{equation}\label{cr}
{{\sum\limits_i {| {{u}({\bm x_i},t + T) - {u}({\bm x_i},t)} |} } \over {\sum\limits_i {| {{u}({\bm x_i},t + T)} |} }} \le 6.0\times10^{-14},
\end{equation}
where $|\cdot|$ denotes the absolute value operator and the summation of $i$ is over the entire system.

We first carried out a set of simulations to get the velocity profiles across the channel at different times. Fig. \ref{eta2.8} shows the velocity profiles in the lines $x= 1, z=0$ at four different times: $t=T/4, T/2, 3T/4$ and $T$. The relaxation time $\tau$ is chosen to be $0.6178$ and $0.5500$ with the Womersley number $\eta=2.8285$ and $4.3416$, respectively. The velocity has been normalized with $U_{max}$, which is the maximal horizontal velocity of Poiseuille flow with pressure gradient $-G$. $U_{max}=1.876\times10^{-2}$ and $4.420\times10^{-2}$ for $\eta=2.8285$ and $4.3416$. The velocity profiles at different $z$ and $x$ locations are also plotted in Fig. \ref{difz} and Fig. \ref{difx}. From above figures, it can be seen that the simulation results agree with the analytical solutions exactly.

Next, we tried to get the accuracy of iD3Q14 MRT model for unsteady pulsatile flow by measuring the convergence order of $GRE_u$.
In the simulations, $\tau$ and $\nu$ are kept unchanged but $c$ increases to two times when $\delta x$ decreases to half, $\eta=2.8285$. The simulation results at the time $t=T/4$, $T/2$, $3T/4$ and $T$ are shown in Table \ref{GREuw}. We also depicted the variation of $GRE_u$ with the lattice spacing in Fig. \ref{edxw}. It can be seen that $-$ln$(GRE_u)$ changes linearly with the $-$ln$(\Delta x)$. The slopes of fitting lines are 1.98, 1.87, 1.98 and 1.86 for the solutions at the time $T/4$, $T/2$, $3T/4$ and $T$, respectively. The above results show that the proposed iD3Q14 MRT model is of second order accuracy in space when simulating unsteady pulsatile flow.

Finally, to compare the present MRT model with d'Humi\`{e}res et al. MRT model, we calculate $GRE_u$ at the time $t=T$ for two MRT models. It should be noted that d'Humi\`{e}res et al. MRT models have two types, one is based on Qian LBGK model and the other one is based on He-Luo LBGK model. Since He-Luo LBGK model has smaller compressible effect, we choose the He-Luo type MRT model for comparison. In the comparison, the numerical results are obtained with the following parameters: $\delta t=0.05$ is fixed, and the grid is $41\times21\times21$, $c$ is equal to 1. $\tau$ is set to be 0.9 while the relaxation parameter $s_{\varepsilon}$ in d'Humi\`{e}res et al. MRT model is set to be 1.0. Table \ref{hlmrt.vs.imrt} shows the variation of $GRE_u$ with the pressure drop along the channel. It can be seen that the $GRE_u$ of the present iD3Q14 MRT model is always smaller than that of d'Humi\`{e}res et al. D3Q15 MRT model, which demonstrates that iD3Q14 MRT model is more accurate than d'Humi\`{e}res et al. D3Q15 MRT model for unsteady pulsatile flow. This may be attributed to that the compressible effect of our proposed model is smaller than that of d'Humi\`{e}res et al. MRT model.

\subsection{3D lid-driven cavity flow}
\label{4.3}
Because the 3D lid-driven cavity flow presents a variety of vortex motions, it is usually used in the validation of numerical method. Ku has simulated this flow with the pseudospectral method \cite{3DCavityFlow}, and the result is widely used as a benchmark for the 3D lid-driven cavity flow.
In this subsection, we also use the result by Ku to validate our method.

The schematic of three-dimensional lid-driven cavity flow is shown in Fig. \ref{cavity-schematic}. We can see that the flow is confined in a cubic box $[0,1]\times[0,1]\times[0,1]$ and driven by the top lid, which is sliding with constant velocity $U_0=1.0$. The flow in the cavity is supposed to be governed by the three-dimensional incompressible N-S equations. The Reynolds number is defined as $U_0L/\nu$, where $L=1.0$ is the length of cubic box, $\nu$ is the kinematic viscosity.

In the simulations, the initial states of velocity and pressure fields, the velocities on the solid walls are all set to be zeros. At the section of $z=0.5$, the symmetric boundary condition is set, i.e.,
\begin{equation}\label{z05symmetry}
\frac{{\partial u}}{{\partial z}} = \frac{{\partial v}}{{\partial z}} = \frac{{\partial p}}{{\partial z}} = 0,{\rm{ }}w = 0.
\end{equation}
At $x=0$ and $x=1$ on $y=1$, $u$ is set to be zero. At the points next to $x=0$ and $x=1$ on $y=1$, $u=0.3$ and $1.0$ are set. The setting of initial and boundary conditions is the same with that of Ku.

In addition, $c=10$ is fixed to make the flow satisfy the low Mach number limit. The convergence criterion for the velocity field is,
\begin{equation}\label{cavity_creterion}
{{\sum\limits_i \sum\limits_k {| {{u_k}({\bm x_i},t + 1000\delta t) - {u_k}({\bm x_i},t)} |} } \over {\sum\limits_i \sum\limits_k {| {{u_k}({\bm x_i},t + 1000\delta t)} |} }} \le 1.0\times10^{-14},
\end{equation}
where the summations of $i$ and $k$ are carried out over all the grid points and the velocities in all directions.

The grid independence of simulation result for iD3Q14 MRT model is firstly examined. The $u_{min}$ in the vertical center line ($z=0.5$ and $x=0.5$) for cavity flow at Re=1000 by using four grids ($49\times49\times25$, $65\times65\times33$, $81\times81\times41$ and $97\times97\times49$) are $-0.2619$, $-0.2693$, $-0.2730$, $-0.2751$, respectively. The deviation of $u_{min}$ between two adjacent grids are $0.0074$, $0.0037$ and $0.0021$. Taking $u_{min}$ at grid $97\times97\times49$ as a benchmark, the relative errors for three smaller grids from $49\times49\times25$ to $81\times81\times41$ are $5.04\%$, $4.24\%$ and $2.83\%$.

Then the simulation results at grid $97\times97\times49$ are used to compare with those of Ku for cavity flow at Re=100, 400 and 1000. The comparison is shown in Fig. \ref{cavityplot}. It can be seen that the agreement between our results and those of Ku are excellent.

Next, the stability of iD3Q14 MRT model is compared with that of iD3Q15 LBGK model. We firstly use the iD3Q15 LBGK model to perform the simulation of cavity flow at Re=1000 with the grid $97\times97\times49$. The horizontal velocity profile in the vertical center line is shown in Fig. \ref{imrt_vs_ibgk}. It can be seen that the result of iD3Q15 LBGK model agree well with the results of iD3Q14 MRT model and Ku.

Since many experimental and numerical studies on 3D lid-driven cavity flow \cite{3DCavityFlow_HouS} are focused on the case of Re=3200, we also choose this case to test the stability of iD3Q14 MRT model. The simulation results of iD3Q14 MRT model and iD3Q15 LBGK model with the grid $97\times97\times49$ are shown in Fig. \ref{yzplane_imrt_vs_ibgk}. In the figure, the velocity vectors on the $yz$ plane at $x=0.5$ and $t=50000\delta t$ are plotted for two models. It should be noted that, in the plotting, the velocities on the left half plane are obtained from those on the right half one by the symmetry transformation, and only one grid point in every two is shown. From Fig. \ref{yzplane_imrt_vs_ibgk}, it can be seen that both models can reproduce the Taylor-G\"{o}rtler-Like (TGL) vortices (two pairs) at the bottom of the plane. It is also found that the flow above the TGL vortices (especially at the two top corners) are different for two models. This may be due to that the instantaneous flow tends to be affected by the model parameters for the unsteady cavity flow at Re=3200.

Finally, we use iD3Q14 MRT model and iD3Q15 LBGK model to simulate the cavity flow at Re=3200 with the grid $49\times49\times25$. The computation is stable for iD3Q14 MRT model, while the computation blows up for iD3Q15 LBGK model. The result of iD3Q14 MRT model is shown in Fig. \ref{cavity_imrt14_49}. The figure shows the velocity vector plot of $yz$ plane at $x=0.5$ and $t=25000\delta t$. It should be noted that $\delta t$ at the current grid is twice that at the grid $97\times97\times49$. From Fig. \ref{cavity_imrt14_49}, it can be seen that two pairs of TGL vortices can still be seen at the bottom of the plane. The above comparison demonstrates that iD3Q14 MRT model is more stable than iD3Q15 LBGK model. This may be attributed to that the iD3Q14 MRT model has more adjustable relaxation parameters.

\section{Conclusions}
\label{conclusions}
We have proposed 14-velocity and 18-velocity multiple-relaxation-time lattice Boltzmann models for three-dimensional incompressible flows and recovered the incompressible N-S equations through the C-E analysis. These two models only use 14 and 18 discrete velocities in velocity space and thus $14\times14$ and $18\times18$ transformation matrixes in moment space, which can reduce the storage and computation costs in simulations. New models are constructed based on the three-dimensional He et al. LBGK models and d'Humi\`{e}res et al. MRT LB models by realizing that He et al. LBGK models only need 14 and 18 discrete velocities without $\bm c_0=(0,0)$, and at the same time, d'Humi\`{e}res et al. MRT LB models contain a moment which has no effect on the recovered macroscopic N-S equations.

We also have carried out the numerical simulations for the 3D steady Poiseuille flow, unsteady pulsatile flow in a square channel and the 3D lid-driven cavity flow using the proposed 14-velocity MRT model. For the Poiseuille flow, we have calculated the horizontal velocity profiles versus $y$ at different $x$ and $z$ sections and the pressure profiles along the channel at different $y$ and $z$ sections. Our numerical results are in excellent agreement with the analytical solutions. As to the pulsatile flow, the horizontal velocity profiles versus $y$ in the center line of the channel ($x=1$ and $z=0$) at four different times $t=T/4$, $T/2$, $3T/4$ and $T$ were computed. The velocity profiles at different $x$ and $z$ sections were also measured. All these numerical results precisely agree with analytical solutions. For the lid-driven cavity flow, the velocity profiles in the vertical and horizontal center lines at section $z=0.5$ were calculated for Re=100, 400 and 1000. The simulation results agree very well with the previous numerical results, which were obtained with an accurate pseudospectral method.

For the steady Poiseuille flow and the unsteady pulsatile flow, we have also conducted the simulations to explore the numerical accuracy of the proposed MRT model. It is found that the proposed model has second-order accuracy in space. We also computed the global relative error in the velocity field of pulsatile flow, versus the pressure drop along the channel for the proposed MRT model and d'Humi\`{e}res et al. MRT model. It is found that the global relative error of our model is always smaller than that of d'Humi\`{e}res et al.. For the cavity flow, we have simulated the flow at Re=3200. It is observed that our MRT model can capture the TGL vortices at the grid $49\times49\times25$, while He et al. LBGK model diverges at this grid.

In short, we have developed two three-dimensional MRT LB models, which can recover to the incompressible N-S equations in the low Mach number limit. These two models have higher storage and computation efficiency than the existing three-dimensional MRT LB models. The new models are based on He et al. LBGK models and d'Humi\`{e}res et al. MRT models, but the new models are more stable or accurate.

\section*{Acknowledgments}
This work is supported by the National Natural Science Foundation of China (Grant Nos 11272132, 51306133 and 11302073).


\begin{appendix}
\section{The Chapman-Enskog analysis for iD3Q14 MRT LB model: derivation of incompressible Navier-Stokes equations}
\label{appendixA}
In this section, we perform the C-E analysis for iD3Q14 MRT LB model to recover the incompressible N-S equations. With the knowledge that in the incompressible flow,
\begin{subequations}
\label{appendix_incompressible_flow_condition}
\begin{equation}
O(\delta p) = O(\delta \rho ) = O({M^2}),
\end{equation}
\begin{equation}
\label{appendix_incompressible_flow_condition2}
O(\bm u) = O(M),
\end{equation}
\end{subequations}
where $M$ represents Mach number, $\delta p$ and $\delta \rho$ are the pressure and density fluctuations, respectively.

We first introduce the following expansions:
\begin{subequations}
\label{aped_firstexpansions}
\begin{equation}
\label{aped_firstexpansions1}
{f_\alpha }(\bm x + {\bm c_\alpha }{\delta _t},t + {\delta _t}) = \sum\limits_{n = 0}^\infty  {\frac{{{\varepsilon ^n}}}{{n!}}} {D_{t}^n}{f_\alpha }(\bm x,t),
\end{equation}
\begin{equation}
\label{aped_firstexpansions2}
{f_\alpha } = \sum\limits_{n = 0}^\infty  {{\varepsilon ^n}f_\alpha ^{(n)}},
\end{equation}
\begin{equation}
\label{aped_firstexpansions3}
{\partial _t} = \sum\limits_{n = 0}^\infty  {{\varepsilon ^n}{\partial _{{t_n}}}},
\end{equation}
\end{subequations}
where $\varepsilon=\delta t$ and ${D_t} \equiv ({\partial _t} + {\bm c_\alpha } \cdot \nabla )$, we can rewrite the lattice Boltzmann equation
\begin{equation}\label{appendix_MRT_evolution_equation}
{f_\alpha }({\bm{x}} + {{\bm{c}}_\alpha }{\delta _t},t + {\delta _t}) - {f_\alpha }({\bm{x}},t) =  - \sum\limits_{i = 1}^{14} {{\Lambda _{\alpha i}}({f_i}({\bm{x}},t) - f_i^{(eq)}({\bm{x}},t))}
\end{equation}
in different orders of $\varepsilon$ as follows:
\begin{subequations}
\label{aped_diforders_eps}
\begin{equation}
\label{aped_diforders_eps0}
O({\varepsilon ^0}): f_\alpha ^{(0)} = f_\alpha ^{(eq)},
\end{equation}
\begin{equation}
\label{aped_diforders_eps1}
O({\varepsilon ^1}): {D_{t_0}}f_\alpha ^{(0)} =  - \sum\limits_{i = 1}^{14} {{{\Lambda}_{\alpha i}}f_i^{(1)}},
\end{equation}
\begin{equation}
\label{aped_diforders_eps2}
O({\varepsilon ^2}): {\partial _{{t_1}}}f_\alpha ^{(0)} + {D_{t_0 }}(f_\alpha ^{(1)} - \frac{1}{2}\sum\limits_{i = 1}^{14} {{\Lambda _{\alpha i}}f_i^{(1)}} ) =  - \sum\limits_{i = 1}^{14} {{{\Lambda}_{\alpha i}}f_i^{(2)}},
\end{equation}
\end{subequations}
where ${D_{t_0}} \equiv {\partial _{{t_0}}} + {\bm c_\alpha } \cdot {\nabla}$, and Eq.(\ref{aped_diforders_eps1}) has been substituted into Eq.(\ref{aped_diforders_eps2}). Then we can transform Eq. (\ref{aped_diforders_eps}) into the moment space:
\begin{subequations}
\label{aped_vectorv2_diforders_eps}
\begin{equation}
\label{aped_vectorv2_diforders_eps0}
{\bm m^{(0)}} = {\bm m^{(eq)}},
\end{equation}
\begin{equation}
\label{aped_vectorv2_diforders_eps1}
({\partial _{{t_0}}}\mathbf I + {\hat \mathbf C_k}{\partial _{k}}){\bm m^{(0)}} =  - \hat \mathbf \Lambda {\bm m^{(1)}},
\end{equation}
\begin{equation}
\label{aped_vectorv2_diforders_eps2}
{\partial _{{t_1}}}{\bm m^{(0)}} + ({\partial _{{t_0}}}\mathbf I + {\hat \mathbf C_k}{\partial _{k}})(\mathbf I - \frac{1}{2}\hat \mathbf \Lambda ){\bm m^{(1)}} =  - \hat \mathbf \Lambda {\bm m^{(2)}},
\end{equation}
\end{subequations}
where ${\hat \mathbf C_k} = \mathbf T{\mathbf C_k}{\mathbf T^{ - 1}}$, $\mathbf{C}_k$ is a diagonal matrix with the $k$ component of every discrete velocity $\bm c_\alpha$ as the element, and
\begin{equation}\label{aped_mi}
{\bm m^{(n)}}= {(0,{e^{(n)}},0,q_x^{(n)},0,q_y^{(n)},0,q_z^{(n)},3p_{xx}^{(n)},p_{\omega \omega }^{(n)},p_{xy}^{(n)},p_{yz}^{(n)},p_{zx}^{(n)},t_{xyz}^{(n)})'}, n=1,2.
\end{equation}
It should be noted that, making the analysis to He et al. iD3Q15 LBGK model, which is similar to the analysis of Guo et al. LBGK model (see Eq. (A15) in Ref. \cite{GuoModel}), we can deduce that $m_1=\sum\limits_{i = 1}^{14} {f_i }(\bm x,t) = (\frac{7}{3}p+\frac{1}{3}u^2)+O(\varepsilon^2+\varepsilon M^2)=m_1^{(eq)}+O(\varepsilon^2+\varepsilon M^2)$.
This means that $P$ is a conserved moment in the low Mach number limit, thus, $P^{(1)}$ and $P^{(2)}$ are set to be zeros in Eq. (\ref{aped_mi}).

Expanding Eq. (\ref{aped_vectorv2_diforders_eps1}), we have
\begin{equation}\label{aped_t0}\tiny
{\partial _{{t_0}}}\left[ {\begin{array}{*{20}{c}}
{7p/3 + u^2/3}\\
{ - 7p + u^2}\\
{{u_x}}\\
{ - 7{u_x}/3}\\
{{u_y}}\\
{ - 7{u_y}/3}\\
{{u_z}}\\
{ - 7{u_z}/3}\\
{3u_x^2 - u^2}\\
{u_y^2 - u_z^2}\\
{{u_x}{u_y}}\\
{{u_y}{u_z}}\\
{{u_z}{u_x}}\\
0
\end{array}} \right] + {\partial _{x}}\left[ {\begin{array}{*{20}{c}}
{{u_x}}\\
{ - 5{u_x}/3}\\
{p + u_x^2}\\
{5u^2/3 - 4u_x^2 - 7p/3}\\
{{u_x}{u_y}}\\
{{u_x}{u_y}}\\
{{u_x}{u_z}}\\
{{u_x}{u_z}}\\
{4{u_x}/3}\\
0\\
{{u_y}/3}\\
0\\
{{u_z}/3}\\
{{u_y}{u_z}}
\end{array}} \right] + {\partial _{y}}\left[ {\begin{array}{*{20}{c}}
{{u_y}}\\
{ - 5{u_y}/3}\\
{{u_x}{u_y}}\\
{{u_x}{u_y}}\\
{p + u_y^2}\\
{5u^2/3 - 4u_y^2 - 7p/3}\\
{{u_y}{u_z}}\\
{{u_y}{u_z}}\\
{ - 2{u_y}/3}\\
{2{u_y}/3}\\
{{u_x}/3}\\
{{u_z}/3}\\
0\\
{{u_x}{u_z}}
\end{array}} \right] + {\partial _{z}}\left[ {\begin{array}{*{20}{c}}
{{u_z}}\\
{ - 5{u_z}/3}\\
{{u_x}{u_z}}\\
{{u_x}{u_z}}\\
{{u_y}{u_z}}\\
{{u_y}{u_z}}\\
{p + u_z^2}\\
{5u^2/3 - 4u_z^2 - 7p/3}\\
{ - 2{u_z}/3}\\
{ - 2{u_z}/3}\\
0\\
{{u_y}/3}\\
{{u_x}/3}\\
{{u_x}{u_y}}
\end{array}} \right] =  - \left[ {\begin{array}{*{20}{c}}
0\\
{{s_e}{e^{(1)}}}\\
0\\
{{s_q}q_x^{(1)}}\\
0\\
{{s_q}q_y^{(1)}}\\
0\\
{{s_q}q_z^{(1)}}\\
{3{s_\nu }p_{xx}^{(1)}}\\
{{s_\nu }p_{\omega \omega }^{(1)}}\\
{{s_\nu }p_{xy}^{(1)}}\\
{{s_\nu }p_{yz}^{(1)}}\\
{{s_\nu }p_{zx}^{(1)}}\\
{{s_t}t_{xyz}^{(1)}}
\end{array}} \right]
\end{equation}
From above equations, we can obtain
\begin{equation}\label{aped_t01}
{\partial _{{t_0}}}(7p/3 + u^2/3)+{\partial _{x}}{u_x} + {\partial _{y}}{u_y} + {\partial _{z}}{u_z} = 0,
\end{equation}
\begin{subequations}\label{aped_t0part}
\begin{equation}\label{aped_t02}
{\partial _{{t_0}}}{u_x} + {\partial _{x}}(p + u_x^2) + {\partial _{y}}({u_x}{u_y}) + {\partial _{z}}({u_x}{u_z}) = 0,
\end{equation}
\begin{equation}\label{aped_t03}
{\partial _{{t_0}}}{u_y} + {\partial _{x}}({u_x}{u_y}) + {\partial _{y}}(p + u_y^2) + {\partial _{z}}({u_y}{u_z}) = 0,
\end{equation}
\begin{equation}\label{aped_t04}
{\partial _{{t_0}}}{u_z} + {\partial _{x}}({u_x}{u_z}) + {\partial _{y}}({u_y}{u_z}) + {\partial _{z}}(p + u_z^2) = 0.
\end{equation}
\end{subequations}
Since $O(\delta p)=O(M^2)$ and $O(\bm u)=O(M)$, we have ${\partial _{{t_0}}}(7p/3 + u^2/3)=O(M^2)$. Omitting the $O(M^2)$ term, Eq. (\ref{aped_t01}) becomes
\begin{equation}\label{appendix_continuity}
{\partial _{x}}{u_x} + {\partial _{y}}{u_y} + {\partial _{z}}{u_z} = 0,
\end{equation}
which is the continuity equation of incompressible N-S equations.

From the expansion of Eq. (\ref{aped_vectorv2_diforders_eps2}), we can have
\begin{subequations}\label{aped_t1part}
\begin{equation}\label{aped_t12}
\begin{array}{l}
{\partial _{{t_1}}}{u_x} + {\partial _{x}}[(1 - {s_\nu }/2)p_{xx}^{(1)} + 2(1 - {s_e}/2){e^{(1)}}/21] + {\partial _{y}}[(1 - {s_\nu }/2)p_{xy}^{(1)}]\\
 + {\partial _{z}}[(1 - {s_\nu }/2)p_{zx}^{(1)}] = 0,
\end{array}
\end{equation}
\begin{equation}\label{aped_t13}
\begin{array}{l}
{\partial _{{t_1}}}{u_y} + {\partial _{x}}[(1 - {s_\nu }/2)p_{xy}^{(1)}] + {\partial _{y}}[(1 - {s_\nu }/2)(p_{\omega \omega }^{(1)} - p_{xx}^{(1)})/2 + 2(1 - {s_e}/2){e^{(1)}}/21]\\
 + {\partial _{z}}[(1 - {s_\nu }/2)p_{yz}^{(1)}] = 0,
\end{array}
\end{equation}
\begin{equation}\label{aped_t14}
\begin{array}{l}
{\partial _{{t_1}}}{u_z} + {\partial _{x}}[(1 - {s_\nu }/2)p_{zx}^{(1)}] + {\partial _{z}}[2(1 - {s_e}/2){e^{(1)}}/21 - (1 - {s_\nu }/2)(p_{\omega \omega }^{(1)} + p_{xx}^{(1)})/2]\\
 + {\partial _{y}}[(1 - {s_\nu }/2)p_{yz}^{(1)}] = 0,
\end{array}
\end{equation}
\end{subequations}
Combining Eq. (\ref{aped_t0part}) and Eq. (\ref{aped_t1part}) with the operation (\ref{aped_t0part})+$\varepsilon\times$(\ref{aped_t1part}), we can obtain
\begin{subequations}\label{aped_tpart}
\begin{equation}\label{aped_t2}
\begin{array}{*{20}{l}}
{{\partial _t}{u_x} + {\partial _x}(p + u_x^2) + {\partial _y}({u_x}{u_y}) + {\partial _z}({u_x}{u_z}) =  - \varepsilon {\partial _y}[(1 - {s_\nu }/2)p_{xy}^{(1)}]}\\
{ - \varepsilon {\partial _x}[(1 - {s_\nu }/2)p_{xx}^{(1)} + 2(1 - {s_e}/2){e^{(1)}}/21] - \varepsilon {\partial _z}[(1 - {s_\nu }/2)p_{zx}^{(1)}],}
\end{array}
\end{equation}
\begin{equation}\label{aped_t3}
\begin{array}{l}
{\partial _t}{u_y} + {\partial _x}({u_x}{u_y}) + {\partial _y}(p + u_y^2) + {\partial _z}({u_y}{u_z}) =  - \varepsilon {\partial _x}[(1 - {s_\nu }/2)p_{xy}^{(1)}]\\
 - \varepsilon {\partial _y}[(1 - {s_\nu }/2)(p_{\omega \omega }^{(1)} - p_{xx}^{(1)})/2 + 2(1 - {s_e}/2){e^{(1)}}/21] - \varepsilon {\partial _z}[(1 - {s_\nu }/2)p_{yz}^{(1)}]{\rm{  }},
\end{array}
\end{equation}
\begin{equation}\label{aped_t4}
\begin{array}{*{20}{l}}
{{\partial _t}{u_z} + {\partial _x}({u_x}{u_z}) + {\partial _y}({u_y}{u_z}) + {\partial _z}(p + u_z^2) =  - \varepsilon {\partial _x}[(1 - {s_\nu }/2)p_{zx}^{(1)}]}\\
{ - \varepsilon {\partial _y}[(1 - {s_\nu }/2)p_{yz}^{(1)}] - \varepsilon {\partial _z}[2(1 - {s_e}/2){e^{(1)}}/21 - (1 - {s_\nu }/2)(p_{\omega \omega }^{(1)} + p_{xx}^{(1)})/2]}.
\end{array}
\end{equation}
\end{subequations}
According to Eq. (\ref{aped_t0}), yields
\begin{subequations}\label{aped_si}
\begin{equation}\label{aped_si1}
{e^{(1)}} =  - \frac{1}{{{{s}_e}}}[{\partial _{{t_0}}}( - 7p + u^2) - \frac{5}{3}({\partial _{x}}{u_x} + {\partial _{y}}{u_y} + {\partial _{z}}{u_z})],
\end{equation}
\begin{equation}\label{aped_si2}
p_{xx}^{(1)} =  - \frac{1}{{3{{s}_\nu }}}[{\partial _{{t_0}}}(3u_x^2 - u^2) + \frac{4}{3}{\partial _{x}}{u_x} - \frac{2}{3}{\partial _{y}}{u_y} - \frac{2}{3}{\partial _{z}}{u_z}],
\end{equation}
\begin{equation}\label{aped_si3}
p_{\omega \omega }^{(1)} =  - \frac{1}{{{{s}_\nu }}}[{\partial _{{t_0}}}(u_y^2 - u_z^2) + \frac{2}{3}{\partial _{y}}{u_y} - \frac{2}{3}{\partial _{z}}{u_z}],
\end{equation}
\begin{equation}\label{aped_si4}
p_{xy}^{(1)} =  - \frac{1}{{{{s}_\nu }}}[{\partial _{{t_0}}}({u_x}{u_y}) + \frac{1}{3}{\partial _{x}}{u_y} + \frac{1}{3}{\partial _{y}}{u_x}],
\end{equation}
\begin{equation}\label{aped_si5}
p_{yz}^{(1)} =  - \frac{1}{{{{s}_\nu }}}[{\partial _{{t_0}}}({u_y}{u_z}) + \frac{1}{3}{\partial _{y}}{u_z} + \frac{1}{3}{\partial _{z}}{u_y}],
\end{equation}
\begin{equation}\label{aped_si6}
p_{zx}^{(1)} =  - \frac{1}{{{{s}_\nu }}}[{\partial _{{t_0}}}({u_z}{u_x}) + \frac{1}{3}{\partial _{x}}{u_z} + \frac{1}{3}{\partial _{z}}{u_x}],
\end{equation}
\end{subequations}
Using Eq. (\ref{aped_t0part}), we can estimate the terms $\partial_{t_0}u_x^2$, $\partial_{t_0}u_y^2$ and $\partial_{t_0}u_z^2$ as
\begin{subequations}\label{appendix_partialt0u2_analysis}
\begin{equation}\label{appendix_partialt0u2_analysis2}
{\partial _{{t_0}}}{u_x^2} = -2u_x ({\partial _{x}}(p + u_x^2) + {\partial _{y}}({u_x}{u_y}) + {\partial _{z}}({u_x}{u_z}) ),
\end{equation}
\begin{equation}\label{appendix_partialt0u2_analysis3}
{\partial _{{t_0}}}{u_y^2} = -2u_y ({\partial _{x}}({u_x}{u_y}) + {\partial _{y}}(p + u_y^2) + {\partial _{z}}({u_y}{u_z})),
\end{equation}
\begin{equation}\label{appendix_partialt0u2_analysis4}
{\partial _{{t_0}}}{u_z^2} = -2u_z ({\partial _{x}}({u_x}{u_z}) + {\partial _{y}}({u_y}{u_z}) + {\partial _{z}}(p + u_z^2)).
\end{equation}
\end{subequations}
From Eq. (\ref{appendix_incompressible_flow_condition}), we can see that $\partial_{t_0}u_x^2$, $\partial_{t_0}u_y^2$ and $\partial_{t_0}u_z^2$ are all in the order of $O(M^3)$. Omitting the $O(M^3)$ terms, Eq. (\ref{aped_si}) becomes
\begin{subequations}\label{aped_sis}
\begin{equation}\label{aped_sis1}
{e^{(1)}} =  - \frac{4}{{3{{s}_e}}}({\partial _{x}}{u_x} + {\partial _{y}}{u_y} + {\partial _{z}}{u_z})=0,
\end{equation}
\begin{equation}\label{aped_sis2}
p_{xx}^{(1)} =  - \frac{2}{{9{{s}_\nu }}}(2{\partial _{x}}{u_x} - {\partial _{y}}{u_y} - {\partial _{z}}{u_z}),
\end{equation}
\begin{equation}\label{aped_sis3}
p_{\omega \omega }^{(1)} =  - \frac{2}{{3{{s}_\nu }}}({\partial _{y}}{u_y} - {\partial _{z}}{u_z}),
\end{equation}
\begin{equation}\label{aped_sis4}
p_{xy}^{(1)} =  - \frac{1}{{3{{s}_\nu }}}({\partial _{x}}{u_y} + {\partial _{y}}{u_x}),
\end{equation}
\begin{equation}\label{aped_sis5}
p_{yz}^{(1)} =  - \frac{1}{{3{{s}_\nu }}}({\partial _{y}}{u_z} + {\partial _{z}}{u_y}),
\end{equation}
\begin{equation}\label{aped_sis6}
p_{zx}^{(1)} =  - \frac{1}{{3{{s}_\nu }}}({\partial _{x}}{u_z} + {\partial _{z}}{u_x}),
\end{equation}
\end{subequations}
where Eq. (\ref{aped_t01}) has been substituted into Eq. (\ref{aped_sis1}) and the term $\partial_{t_0}u^2$ has been omitted.

Substituting Eq. (\ref{aped_sis}) into Eq. (\ref{aped_tpart}), and supposing $\nu  = c_s^2( \tau- 1/2){\delta t}$, where $\tau=1/{s_\nu }$ and $c_s^2 = 1/3$, then we have
\begin{subequations}\label{aped_macroeq-simple}
\begin{equation}\label{aped_macroeq-simple2}
\begin{array}{l}
{\partial _t}{u_x} + {\partial _x}(u_x^2) + {\partial _y}({u_x}{u_y}) + {\partial _z}({u_x}{u_z})=-{\partial _x}p+\nu (\partial _x^2{u_x} + \partial _y^2{u_x} + \partial _z^2{u_x}),
\end{array}
\end{equation}
\begin{equation}\label{aped_macroeq-simple3}
\begin{array}{l}
{\partial _t}{u_y} + {\partial _x}({u_x}{u_y}) + {\partial _y}(u_y^2) + {\partial _z}({u_y}{u_z})=-{\partial _y}p+\nu (\partial _x^2{u_y} + \partial _y^2{u_y} + \partial _z^2{u_y}),
\end{array}
\end{equation}
\begin{equation}\label{aped_macroeq-simple4}
\begin{array}{l}
{\partial _t}{u_z} + {\partial _x}({u_x}{u_z}) + {\partial _y}({u_y}{u_z}) + {\partial _z}(u_z^2)=-{\partial _z}p+\nu (\partial _x^2{u_z} + \partial _y^2{u_z} + \partial _z^2{u_z}),
\end{array}
\end{equation}
\end{subequations}
which are the momentum equations of incompressible N-S equations. It should be noted that the continuity equation and $\varepsilon=\delta t$ have been used in deriving the above equations.

All in all, from the Chapman-Enskog analysis for iD3Q14 MRT LB model, we can see that, in the low Mach number limit, the proposed 14-velocity model can recover to the incompressible N-S equations, which can be written in a vector form as Eq. (\ref{ns}).

\section{iD3Q18 MRT LB model}
\label{appendixB}
The three-dimensional incompressible MRT LB model with 18-velocity adopts the following discrete velocity directions:
 \begin{equation} \begin{array}{c}
\{\emph{\textbf{c}}_{1}, \emph{\textbf{c}}_{2},\ldots,
\emph{\textbf{c}}_{18} \} =\nonumber
\\ \left\{
\begin{array}{cccccccccccccccccc}
1 & -1 & 0 & 0 & 0 & 0 & 1 & -1 & 1 & -1 & 1 & -1 & 1 & -1 & 0 & 0 & 0 & 0 \\
0 & 0 & 1 & -1 & 0 & 0 & 1 &  1 & -1 & -1 & 0 & 0 & 0 & 0 & 1 & -1 & 1 & -1  \\
0 & 0 & 0 & 0 & 1 & -1 & 0 & 0 & 0 & 0 & 1 & 1 & -1 & -1 & 1&1 &-1 & -1
\end{array} \right\}c,
\nonumber \\ \nonumber \\ \omega_{1-6}=1/18,\
\omega_{7-18}=1/36;\ c_{s}^{2}=c^{2}/3, \ c=1. \nonumber
\end{array}
\label{D3Q19}
\end{equation}
 18 orthogonal basis vectors can be derived by the Gram-Schmidt orthogonalization procedure:
\begin{subequations}
\label{appendix_m18}
\begin{equation}
\left.
\begin{array}{c}
  |\phi_1\rangle_{\alpha}=\left  \| \textbf{c}_\alpha \right \|^0, \\
  |\phi_2\rangle_{\alpha}=3\left  \| \textbf{c}_\alpha \right \|^2-5\left  \| \textbf{c}_\alpha
  \right \|^0,
\end{array}
\right \}
\end{equation}
\begin{equation}
\left.
\begin{array}{c}
  |\phi_3\rangle_{\alpha}=c_{\alpha x}, \\
  |\phi_5\rangle_{\alpha}=c_{\alpha y}, \\
  |\phi_7\rangle_{\alpha}=c_{\alpha z}, \\
\end{array}
\right \}
\end{equation}
\begin{equation}
\left.
\begin{array}{c}
  |\phi_4\rangle_{\alpha}=(5\left  \| \textbf{c}_\alpha \right \|^2-9)c_{\alpha x}, \\
  |\phi_6\rangle_{\alpha}=(5\left  \| \textbf{c}_\alpha \right \|^2-9)c_{\alpha y}, \\
  |\phi_8\rangle_{\alpha}=(5\left  \| \textbf{c}_\alpha \right \|^2-9)c_{\alpha z}, \\
\end{array}
\right \}
\end{equation}
\begin{equation}
\left.
\begin{array}{c}
  |\phi_9\rangle_{\alpha}=3c_{\alpha x}^{2}-\left  \| \textbf{c}_\alpha \right \|^2, \\
  |\phi_{11}\rangle_{\alpha}=c_{\alpha y}^{2}-c_{\alpha z}^{2}, \\
\end{array}
\right \}
\end{equation}
\begin{equation}
\left.
\begin{array}{c}
  |\phi_{13}\rangle_{\alpha}=c_{\alpha x}c_{\alpha y}, \\
  |\phi_{14}\rangle_{\alpha}=c_{\alpha y}c_{\alpha z}, \\
  |\phi_{15}\rangle_{\alpha}=c_{\alpha x}c_{\alpha z}, \\
\end{array}
\right \}
\end{equation}
\begin{equation}
\left.
\begin{array}{c}
  |\phi_{10}\rangle_{\alpha}=(3\left \| \bm c_\alpha\right\|^2-5)(3c_{\alpha x}^2-\left\|\bm c_\alpha\right|^2), \\
  |\phi_{12}\rangle_{\alpha}=(3\left \| \bm c_\alpha\right\|^2-5)(c_{\alpha y}^2-c_{\alpha z}^2), \\
\end{array}
\right \}
\end{equation}
\begin{equation}
\left.
\begin{array}{c}
  |\phi_{16}\rangle_{\alpha}=(c_{\alpha y}^2-c_{\alpha z}^2)c_{\alpha x},\\
  |\phi_{17}\rangle_{\alpha}=(c_{\alpha z}^2-c_{\alpha x}^2)c_{\alpha y}, \\
  |\phi_{18}\rangle_{\alpha}=(c_{\alpha x}^2-c_{\alpha y}^2)c_{\alpha z}, \\
\end{array}
\right \}
\end{equation}
\end{subequations}
 where $\alpha\in\{1,2,\cdots,18\}$. The corresponding 18 moments $\{m_{\beta}(\bm x,t)|\beta=1,2,\cdots,18\}$ are defined as:
 \begin{equation}\label{d3q18moment}
\bm m(\bm{x},t) = (P,e,j_x,q_x,j_y,q_y,j_z,q_z,3p_{xx},3\pi_{xx},p_{\omega \omega},\pi_{\omega \omega},p_{xy},p_{yz},p_{xz},t_{x},t_{y},t_{z})'.
\end{equation}
The diagonal collision matrix is
\begin{equation}\label{sD3Q19}
\bm {\hat  \Lambda}\equiv diag(s_c, s_{e}, s_c, s_{q}, s_c, s_{q}, s_c, s_{q},
s_{\nu}, s_{\pi}, s_{\nu}, s_{\pi}, s_{\nu}, s_{\nu}, s_{\nu},
s_{t}, s_{t}, s_{t}),
\end{equation}
and the transformation matrix $\mathbf{T}$ can be obtained from Eq. (\ref{appendix_m18}),
{\scriptsize
\begin{equation}
\mathbf{T}=\left( {\begin{array}{*{20}c}
   1 & 1 & 1 & 1 & 1 & 1 & 1 & 1 & 1 & 1 & 1 & 1 & 1 & 1 & 1 & 1 & 1 & 1  \\
   { - 2} & { - 2} & { - 2} & { - 2} & { - 2} & { - 2} & 1 & 1 & 1 & 1 & 1 & 1 & 1 & 1 & 1 & 1 & 1 & 1  \\
   1 & { - 1} & 0 & 0 & 0 & 0 & 1 & { - 1} & 1 & { - 1} & 1 & { - 1} & 1 & { - 1} & 0 & 0 & 0 & 0  \\
   { - 4} & 4 & 0 & 0 & 0 & 0 & 1 & { - 1} & 1 & { - 1} & 1 & { - 1} & 1 & { - 1} & 0 & 0 & 0 & 0  \\
   0 & 0 & 1 & { - 1} & 0 & 0 & 1 & 1 & { - 1} & { - 1} & 0 & 0 & 0 & 0 & 1 & { - 1} & 1 & { - 1}  \\
   0 & 0 & { - 4} & 4 & 0 & 0 & 1 & 1 & { - 1} & { - 1} & 0 & 0 & 0 & 0 & 1 & { - 1} & 1 & { - 1}  \\
   0 & 0 & 0 & 0 & 1 & { - 1} & 0 & 0 & 0 & 0 & 1 & 1 & { - 1} & { - 1} & 1 & 1 & { - 1} & { - 1}  \\
   0 & 0 & 0 & 0 & { - 4} & 4 & 0 & 0 & 0 & 0 & 1 & 1 & { - 1} & { - 1} & 1 & 1 & { - 1} & { - 1}  \\
   2 & 2 & { - 1} & { - 1} & { - 1} & { - 1} & 1 & 1 & 1 & 1 & 1 & 1 & 1 & 1 & { - 2} & { - 2} & { - 2} & { - 2}  \\
   { - 4} & { - 4} & 2 & 2 & 2 & 2 & 1 & 1 & 1 & 1 & 1 & 1 & 1 & 1 & { - 2} & { - 2} & { - 2} & { - 2}  \\
   0 & 0 & 1 & 1 & { - 1} & { - 1} & 1 & 1 & 1 & 1 & { - 1} & { - 1} & { - 1} & { - 1} & 0 & 0 & 0 & 0  \\
   0 & 0 & { - 2} & { - 2} & 2 & 2 & 1 & 1 & 1 & 1 & { - 1} & { - 1} & { - 1} & { - 1} & 0 & 0 & 0 & 0  \\
   0 & 0 & 0 & 0 & 0 & 0 & 1 & { - 1} & { - 1} & 1 & 0 & 0 & 0 & 0 & 0 & 0 & 0 & 0  \\
   0 & 0 & 0 & 0 & 0 & 0 & 0 & 0 & 0 & 0 & 0 & 0 & 0 & 0 & 1 & { - 1} & { - 1} & 1  \\
   0 & 0 & 0 & 0 & 0 & 0 & 0 & 0 & 0 & 0 & 1 & { - 1} & { - 1} & 1 & 0 & 0 & 0 & 0  \\
   0 & 0 & 0 & 0 & 0 & 0 & 1 & { - 1} & 1 & { - 1} & { - 1} & 1 & { - 1} & 1 & 0 & 0 & 0 & 0  \\
   0 & 0 & 0 & 0 & 0 & 0 & { - 1} & { - 1} & 1 & 1 & 0 & 0 & 0 & 0 & 1 & { - 1} & 1 & { - 1}  \\
   0 & 0 & 0 & 0 & 0 & 0 & 0 & 0 & 0 & 0 & 1 & 1 & { - 1} & { - 1} & { - 1} & { - 1} & 1 & 1  \\
 \end{array} } \right).
\end{equation}
}
The equilibria of 18 moments are defined as
\begin{subequations}
\label{meq}
\begin{equation}
\left.
\begin{array}{c}
   P^{(eq)}=2p+\frac{1}{2}u^2, \\
   e^{(eq)}=-p+\frac{1}{2}u^2,
\end{array}
\right \}
\end{equation}
\begin{equation}
\left.
\begin{array}{c}
  j_x^{(eq)}=u_x, \\
  j_y^{(eq)}=u_y, \\
  j_z^{(eq)}=u_z, \\
\end{array}
\right \}
\end{equation}
\begin{equation}
\left.
\begin{array}{c}
  q_x^{(eq)}=-\frac{2}{3}u_x, \\
  q_y^{(eq)}=-\frac{2}{3}u_y, \\
  q_z^{(eq)}=-\frac{2}{3}u_z, \\
\end{array}
\right \}
\end{equation}
\begin{equation}
\left.
\begin{array}{c}
  3p_{xx}^{(eq)}=3u_x^2-u^2, \\
  p_{\omega \omega}^{(eq)}=u_{y}^2-u_{z}^2, \\
\end{array}
\right \}
\end{equation}
\begin{equation}
\left.
\begin{array}{c}
  3\pi_{xx}^{(eq)}=-\frac{1}{2}(3u_x^2-u^2), \\
  \pi_{\omega \omega}^{(eq)}=-\frac{1}{2}(u_{y}^2-u_{z}^2), \\
\end{array}
\right \}
\end{equation}
\begin{equation}
\left.
\begin{array}{c}
  p_{xy}^{(eq)}=u_x u_y, \\
  p_{yz}^{(eq)}=u_y u_z, \\
  p_{xz}^{(eq)}=u_x u_z, \\
\end{array}
\right \}
\end{equation}
\begin{equation}
\left.
\begin{array}{c}
  t_{x}^{(eq)}=0, \\
  t_{y}^{(eq)}=0, \\
  t_{z}^{(eq)}=0. \\
\end{array}
\right \}
\end{equation}
\end{subequations}

Through the C-E expansion, the incompressible N-S equations can be recovered in the low Mach number limit for iD3Q18 MRT model. The relationship between the kinematic viscosity and the relaxation parameter $s_{\nu}$ is the same with that of iD3Q14 MRT model. ID3Q18 MRT model can recover to the corresponding He et al. iD3Q19 LBGK model by setting all the relaxation parameters to be $1/\tau$, where $\tau$ is the relaxation time of iD3Q19 LBGK model.

It should be noted that, we also simulated the steady Poiseuille flow, unsteady pulsatile flow and lid-driven cavity flow in three dimensions with iD3Q18 MRT model. The numerical results show that iD3Q18 MRT model is also of second order accuracy in space for steady Poiseuille flow and unsteady pulsatile flow, but iD3Q18 MRT model is more stable than iD3Q14 MRT model for the lid-driven cavity flow. In the simulation of lid-driven cavity flow with the grid $49\times49\times25$ , the reached Reynolds number for iD3Q18 MRT model is larger than that for iD3Q14 MRT model. This finding is consistent with the observation that D3Q19 MRT model is more stable than D3Q15 MRT model by d'Humi\`{e}res et al. \cite{Humieres3DMRT}. In addition, based on our construction method of $(q-1)\times(q-1)$ transformation matrix in three dimensions, we can also obtain iD3Q26 MRT model for three-dimensional incompressible flows. A detailed comparative study of above three MRT models is left for future work.

\end{appendix}





\newpage

Figure \ref{poiseuille-schematic}: The schematic of three-dimensional Poiseuille flow.

Figure \ref{plotmu}: The left figures show the horizontal velocity profiles for Poiseuille flow at section $x=1$, while the right figures show the pressure profiles at section $z=0$. lines: analytical solutions, symbols: numerical results.

Figure \ref{plotp}: The velocity and pressure profiles at different $x$ and $z$ sections for Poiseuille flow.

Figure \ref{edx}: The variation of $GRE_u$ with the lattice spacing for Poiseuille flow by using different $1/\tau$.

Figure \ref{eta2.8}: The variation of $u$ with $y$ for pulsatile flow at the location $x=1,z=0$. lines: analytical solutions, symbols: numerical solutions.

Figure \ref{difz}: The variation of $u$ with $y$ for pulsatile flow at different $z$ locations at section $x=1$ for $\eta=2.8285$.

Figure \ref{difx}: The variation of $u$ with $y$ for pulsatile flow at different $x$ locations at section $z=0$ for $\eta=2.8285$.

Figure \ref{edxw}: The variation of $GRE_u$ with lattice spacing for pulsatile flow at different times, $\eta=2.8285$.

Figure \ref{cavity-schematic}: The schematic of three-dimensional lid-driven cavity flow.

Figure \ref{cav_grid_independence}: The distribution of $u$ at four different grids in the vertical center line ($z=0.5$ and $x=0.5$) for cavity flow at Re=1000, (b) is the magnification of square area in (a).

Figure \ref{cavityplot}: The velocity distribution in the vertical and horizontal center lines at section $z=0.5$ for cavity flow at different Re, $\square$ : the results of Ku, -- : present simulation.

Figure \ref{imrt_vs_ibgk}: The distribution of $u$ in vertical center line for cavity flow at Re=1000. The simulations are carried out with grid $97\times97\times49$.

Figure \ref{yzplane_imrt_vs_ibgk}: The velocity vector plot of $yz$ plane at $x=0.5$ and $t=50000\delta t$ for cavity flow at Re=3200, the grid is $97\times97\times49$. The length of the arrows is three times the actual velocity magnitude.

Figure \ref{cavity_imrt14_49}: The velocity vector on the $yz$ plane at $x=0.5$ and $t=25000\delta t$ for cavity flow at Re=3200, the grid is $49\times49\times25$. The length of the arrows is three times the actual velocity magnitude. The result is obtained from iD3Q14 MRT model, while the simulation by iD3Q15 LBGK model diverges.

\newpage

\setcounter{table}{0}
\renewcommand{\thetable}{\arabic{table}}
\begin{table}
\caption{The $GRE_u$ of Poiseuille flow for different lattice spacings and different $1/\tau$.}
\centering
\begin{tabular}{ccccc}
  \hline \hline
           &                        &   \multicolumn{2}{c}{$GRE_{u}$}                  &                       \\
  \cline{2-5}
  $1/\tau$ & $\Delta x=1/8$         & $\Delta x=1/16$        & $\Delta x=1/32$         & $\Delta x=1/64$        \\
  \hline
  0.8      & $1.116\times10^{-1}$   & $3.277\times10^{-2}$   & $9.001\times10^{-3}$    & $2.371\times10^{-3}$   \\
  \hline
  1.0      & $2.976\times10^{-2}$   & $7.400\times10^{-3}$   & $1.846\times10^{-3}$    & $4.610\times10^{-4}$    \\
  \hline
  1.3      & $5.824\times10^{-2}$   & $1.854\times10^{-2}$   & $5.232\times10^{-3}$    & $1.392\times10^{-3}$    \\
  \hline \hline
\end{tabular}
\label{GREu}
\end{table}

\begin{table}
\caption{The $GRE_u$ of pulsatile flow for different lattice spacings and different times.}
\centering
\begin{tabular}{ccccc}
  \hline \hline
             &                        & \multicolumn{2}{c}{$GRE_{u}$}                  &                        \\
  \cline{2-5}
  $t$        & $\Delta x=1/20$        & $\Delta x=1/40$        & $\Delta x=1/60$       & $\Delta x=1/80$         \\
  \hline
  $T/4$      & $1.750\times10^{-2}$   & $4.465\times10^{-3}$   & $1.994\times10^{-3}$  & $1.124\times10^{-3}$    \\
  \hline
  $T/2$      & $4.060\times10^{-2}$   & $1.169\times10^{-2}$   & $5.444\times10^{-3}$  & $3.134\times10^{-3}$    \\
  \hline
  $3T/4$     & $2.201\times10^{-2}$   & $5.661\times10^{-3}$   & $2.535\times10^{-3}$  & $1.430\times10^{-3}$    \\
  \hline
  $T$        & $3.811\times10^{-2}$   & $1.108\times10^{-2}$   & $5.173\times10^{-3}$  & $2.981\times10^{-3}$    \\
  \hline \hline
\end{tabular}
\label{GREuw}
\end{table}

\begin{table}
\caption{The variation of $GRE_u$ with the pressure drop for pulsatile flow by using different MRT models, $t=T$.}
\begin{tabular}{|c|c|c|c|c|c|}
\hline
\multirow{2}*{$\Delta p$} & \multirow{2}*{$Re$} & \multirow{2}*{$U_{max}$} & \multirow{2}*{$Ma_{max}$} & \multicolumn{2}{c|}{$GRE_u$}
                                                                                            \\\cline{5-6}
                       &                     &                        &                     & {d'Humi\`{e}res et al. D3Q15 MRT} & {iD3Q14 MRT}
\\\hline {0.001} & {0.41} & {0.0055} & {0.0096}  & {0.0067} & {0.0059}
\\\hline {0.005} & {2.1} & {0.028} & {0.047}  & {0.0078} & {0.0067}
\\\hline {0.01} & {4.1} & {0.055} &  {0.096}  & {0.0092} & {0.0075}
\\\hline {0.02} & {8.3} & {0.11} & {0.19}  & {0.012} &  {0.0090}
\\\hline {0.05} & {20.7} & {0.28} & {0.48} & {0.020} & {0.010}
\\\hline
\end{tabular} \label{hlmrt.vs.imrt}
\end{table}


\newpage

\setcounter{figure}{0}
\renewcommand{\thefigure}{\arabic{figure}}

\begin{figure}
\centering
\includegraphics[width=4.0in]{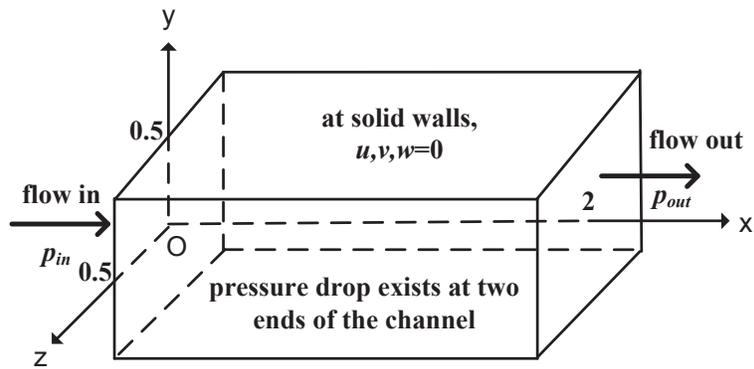}
\caption{The schematic of three-dimensional Poiseuille flow.}
\label{poiseuille-schematic}
\end{figure}

\begin{figure}
\centering (a) $1/\tau=1.2$\\
\includegraphics[width=2.3in]{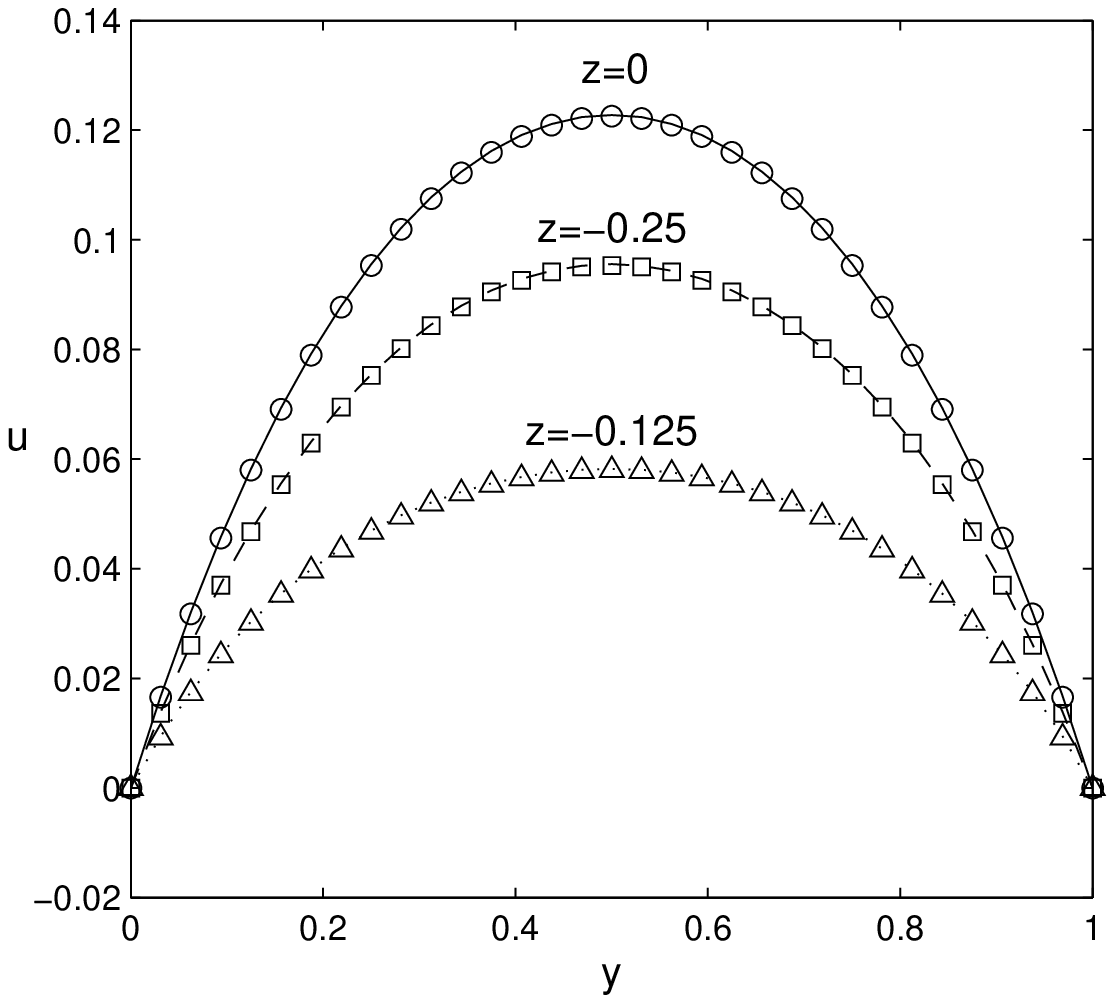}
\includegraphics[width=2.3in]{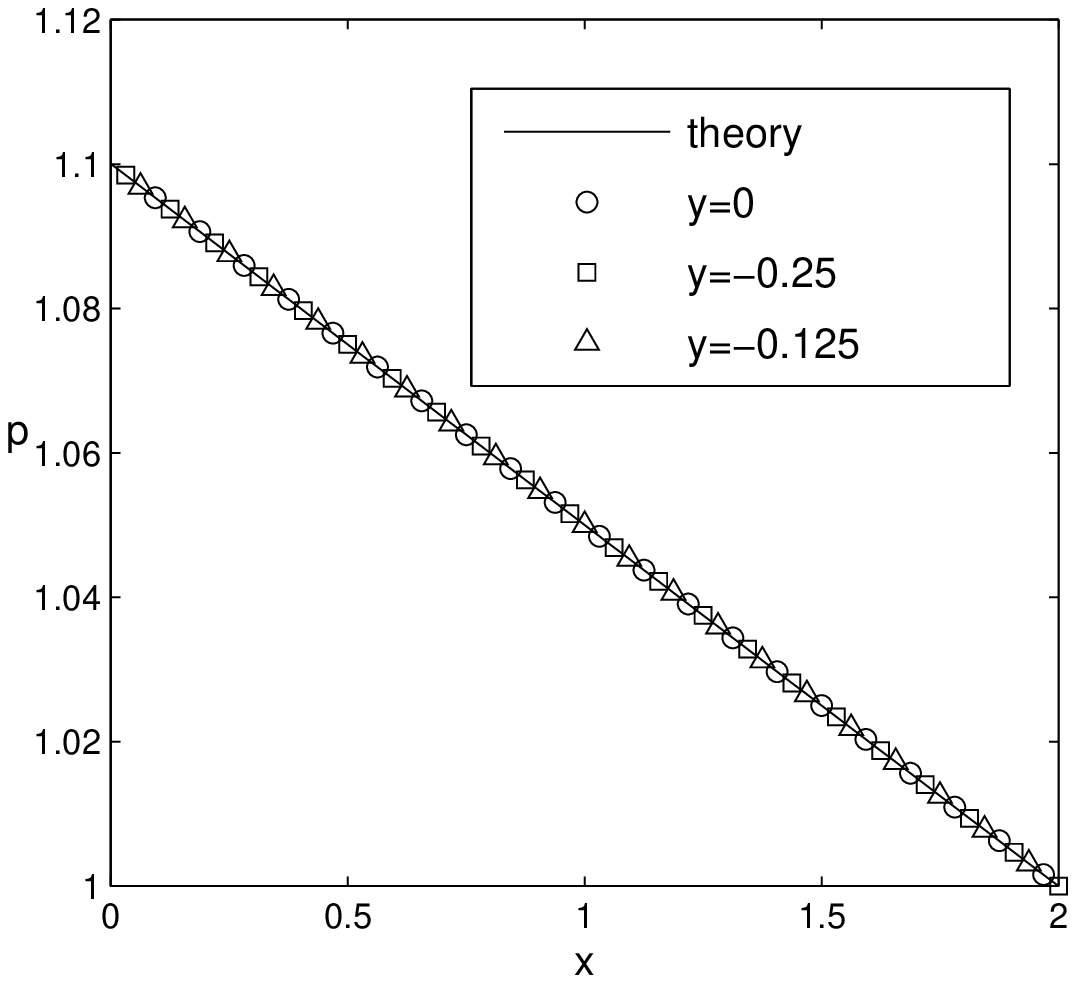}\\
(b) $1/\tau=1.0$\\
\includegraphics[width=2.3in]{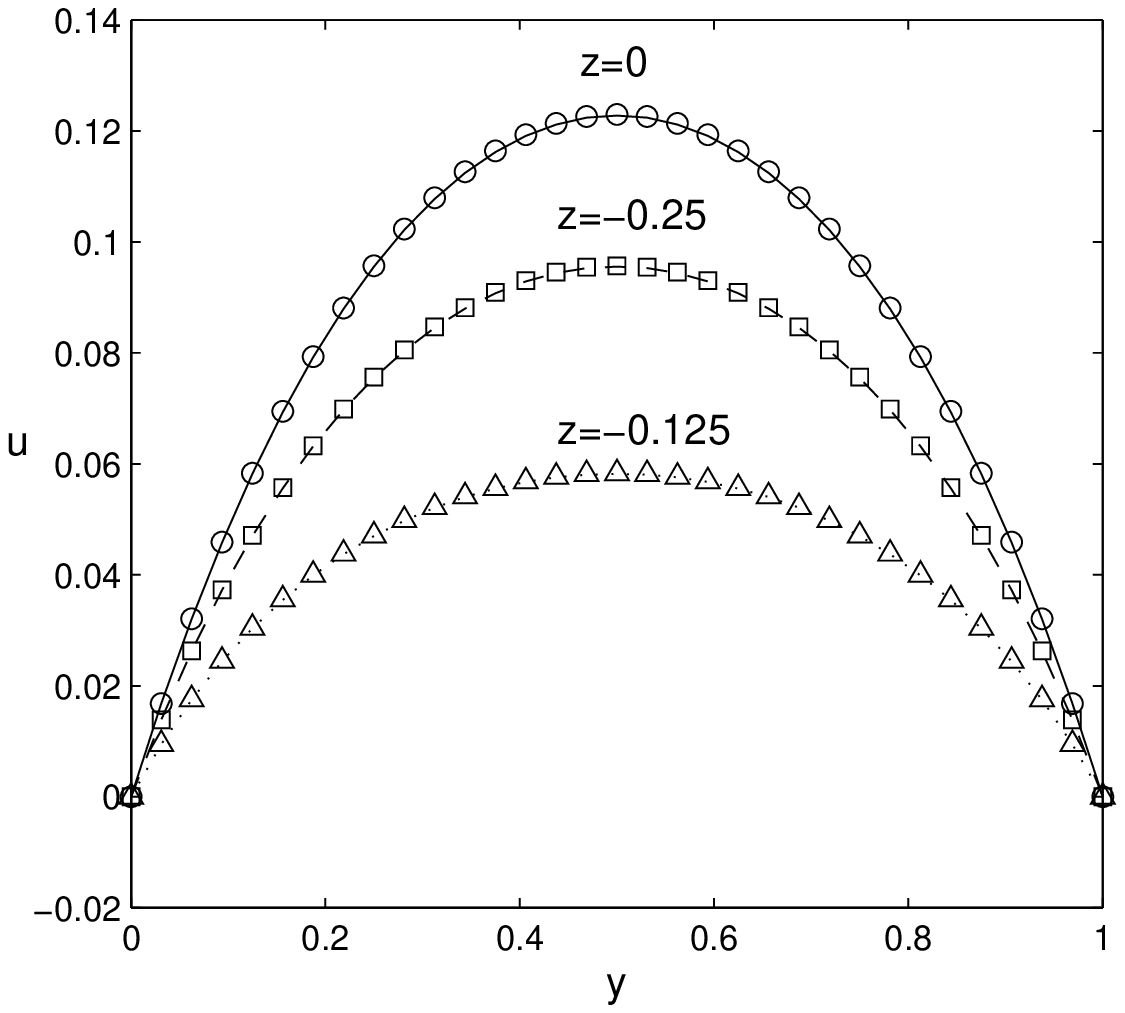}
\includegraphics[width=2.3in]{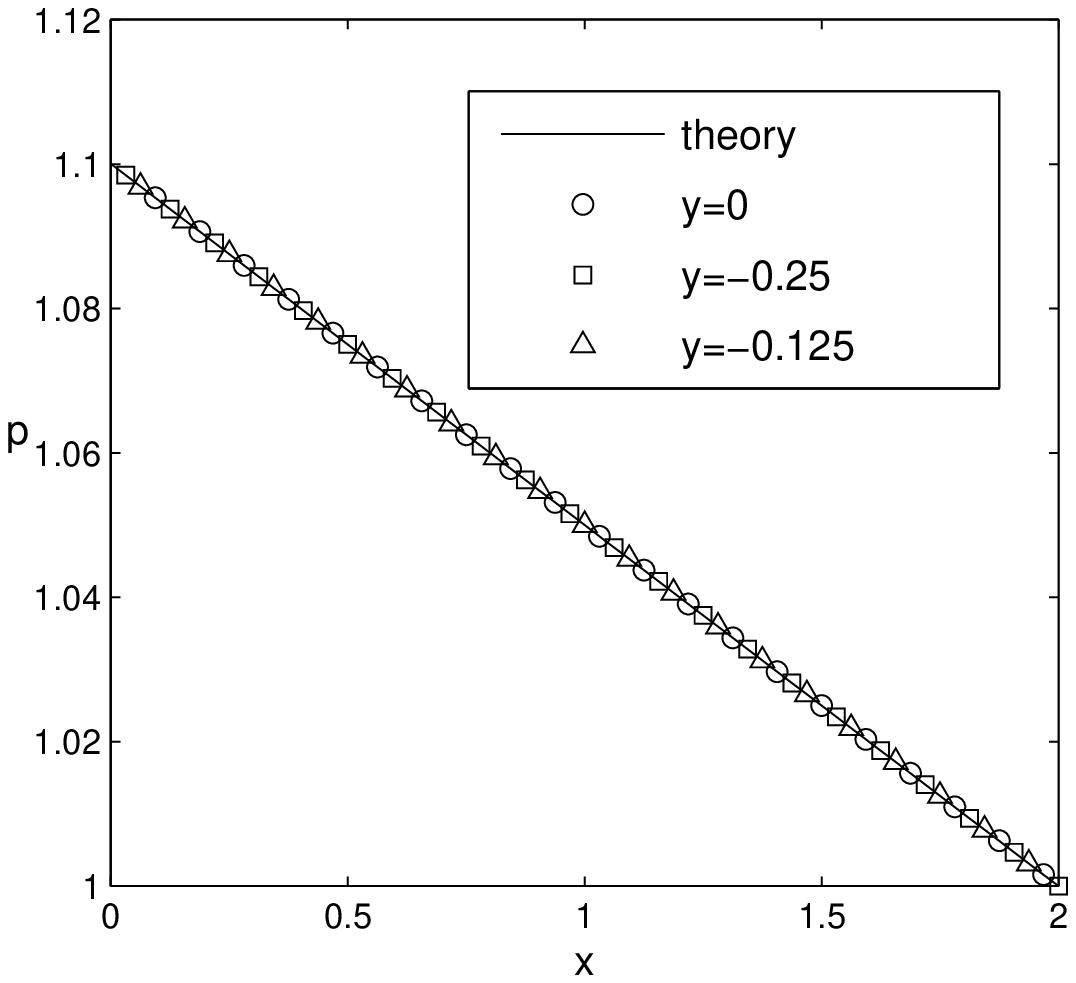}\\
(c) $1/\tau=0.8$\\
\includegraphics[width=2.3in]{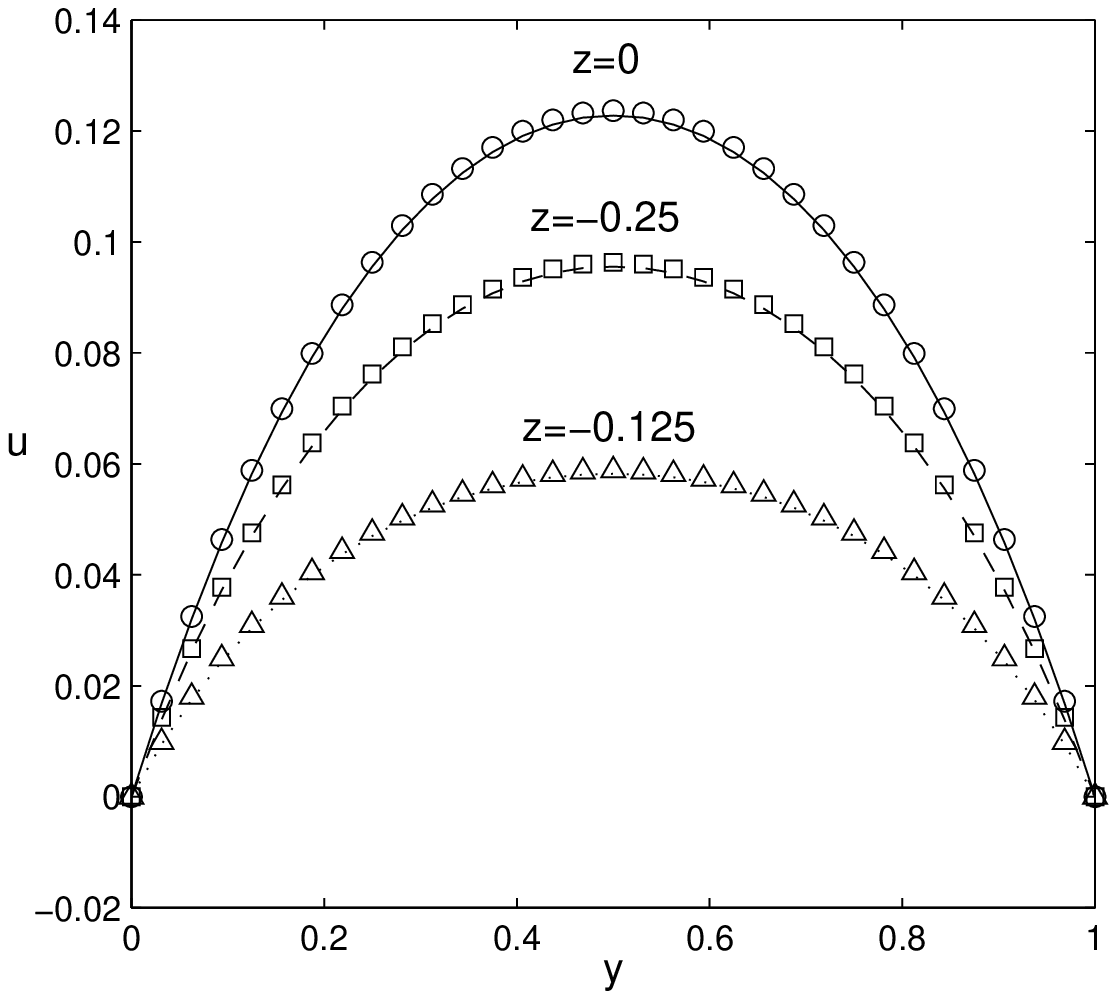}
\includegraphics[width=2.3in]{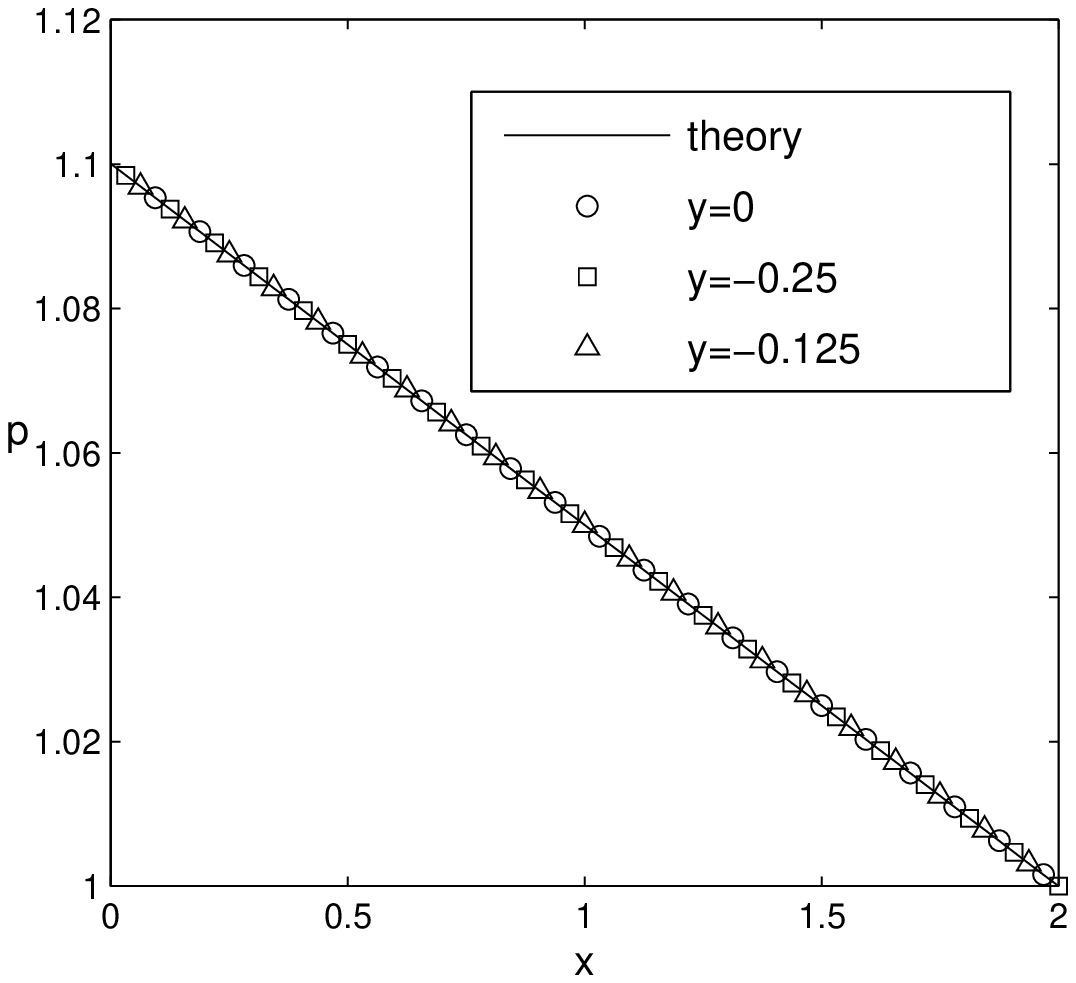}\\
\caption{The left figures show the horizontal velocity profiles for Poiseuille flow at section $x=1$, while the right figures show the pressure profiles at section $z=0$. lines: analytical solutions, symbols: numerical results.}
\label{plotmu}
\end{figure}

\begin{figure}
\centering (a) horizontal velocity profiles at different $x$ sections\\
\includegraphics[width=3.5in]{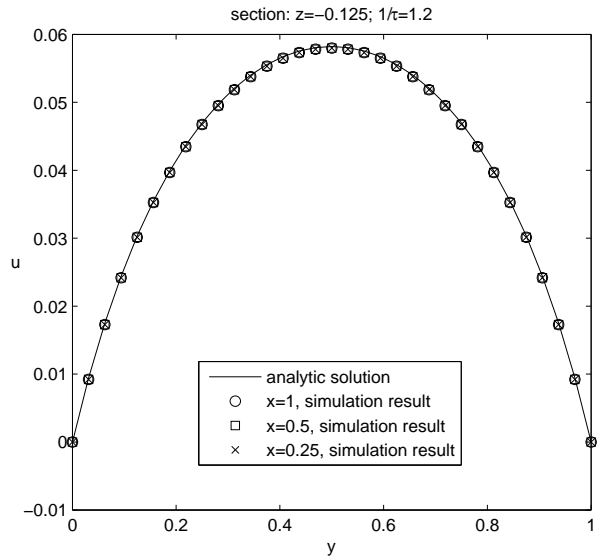}\\
(b) pressure profiles at different $z$ sections\\
\includegraphics[width=3.5in]{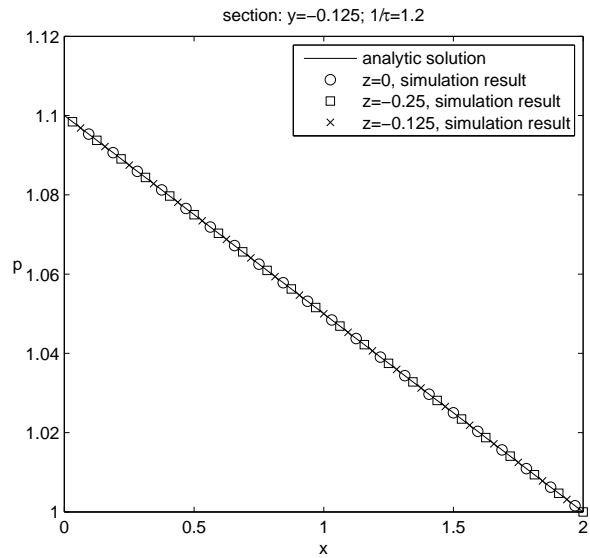}
\caption{The velocity and pressure profiles at different $x$ and $z$ sections for Poiseuille flow.}
\label{plotp}
\end{figure}

\begin{figure}
\centering
\includegraphics[width=4.0in]{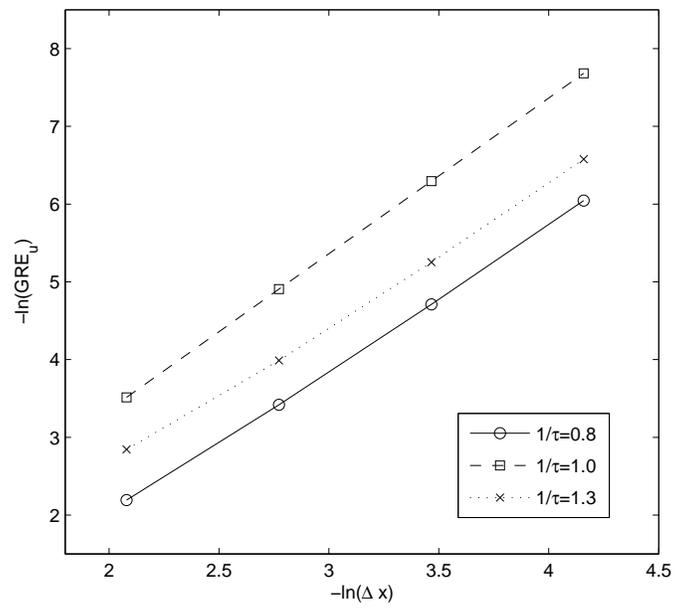}
\caption{The variation of $GRE_u$ with the lattice spacing at different $1/\tau$ for Poiseuille flow.}
\label{edx}
\end{figure}

\begin{figure}
\centering (a) $\eta=2.8285$\\
\includegraphics[width=3.5in]{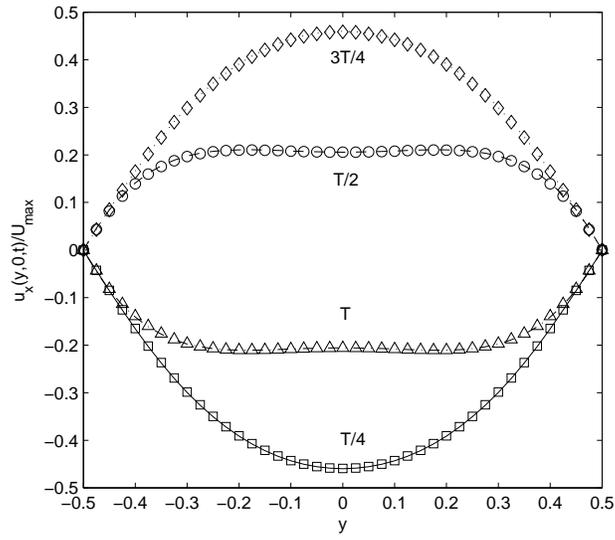}\\
(b) $\eta=4.3416$\\
\includegraphics[width=3.5in]{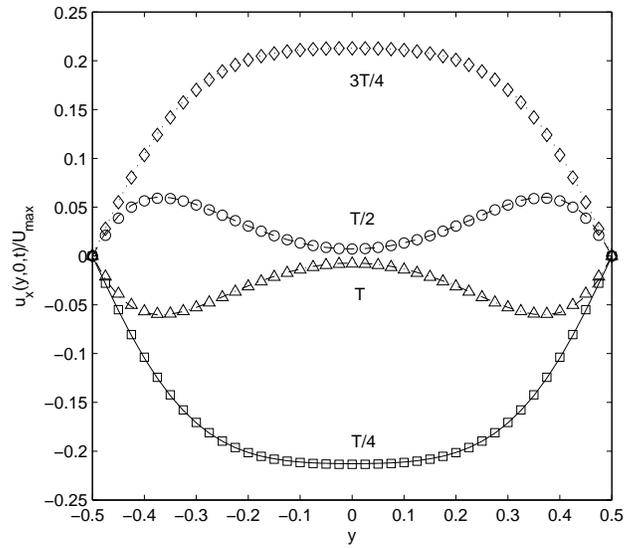}\\
\caption{The variation of $u$ with $y$ for pulsatile flow at the location $x=1,z=0$. lines: analytical solutions, symbols: numerical solutions.}
\label{eta2.8}
\end{figure}

\begin{figure}
\centering
\includegraphics[width=4.0in]{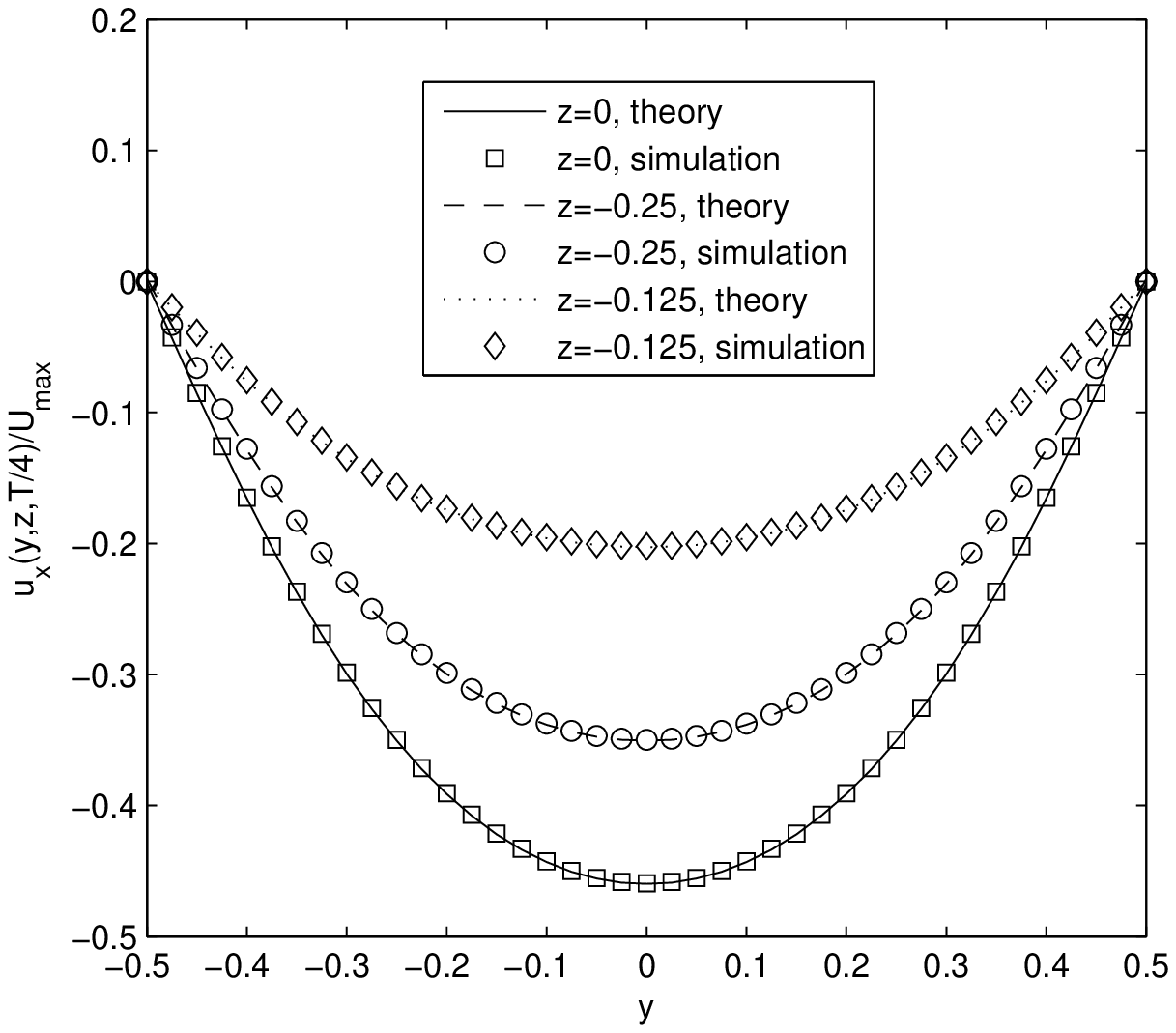}
\caption{The variation of $u$ with $y$ for pulsatile flow at different $z$ locations at section $x=1$ for $\eta=2.8285$.}
\label{difz}
\end{figure}

\begin{figure}
\centering
\includegraphics[width=4.0in]{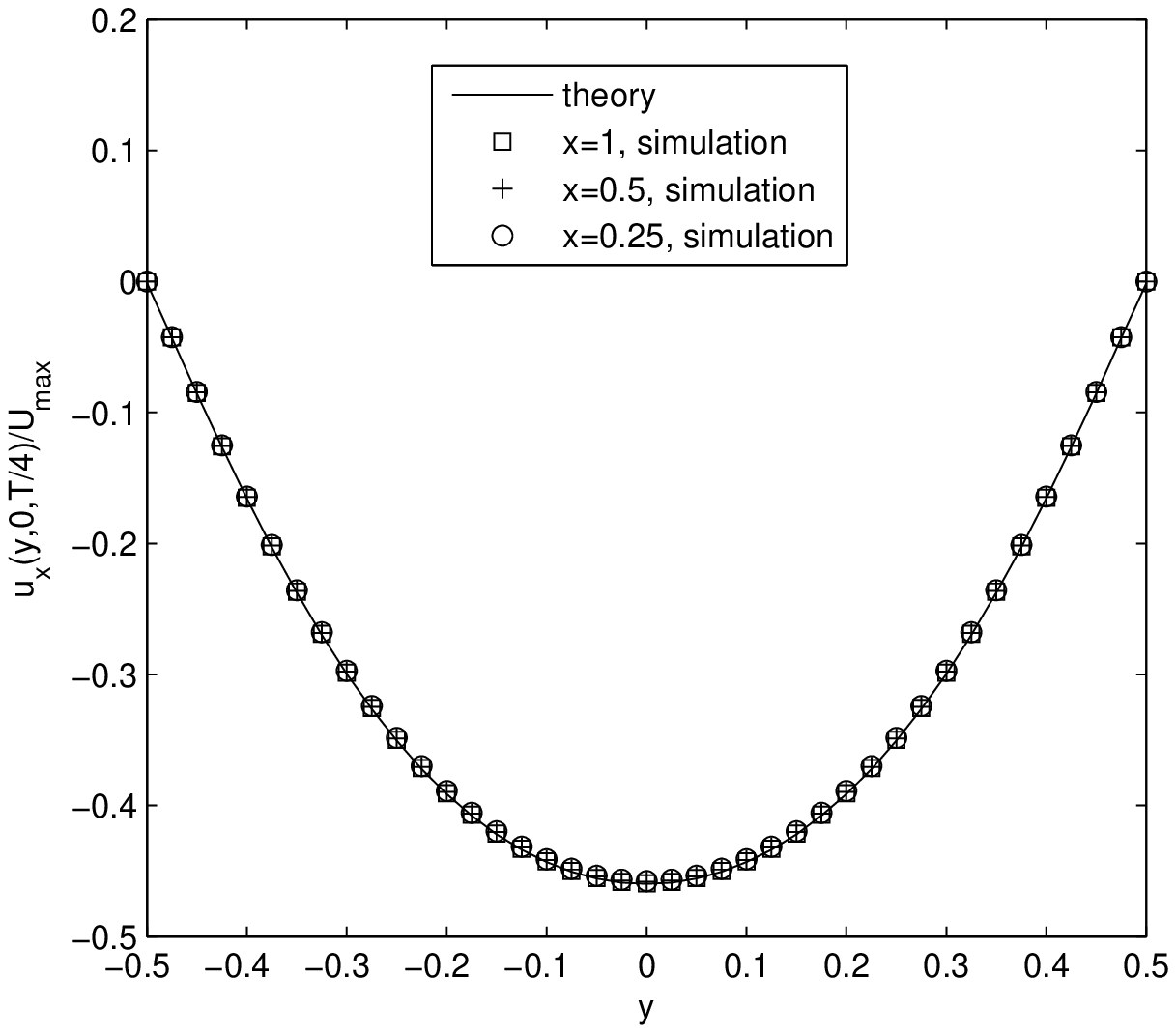}
\caption{The variation of $u$ with $y$ for pulsatile flow at different $x$ locations at section $z=0$ for $\eta=2.8285$.}
\label{difx}
\end{figure}

\begin{figure}
\centering
\includegraphics[width=4.0in]{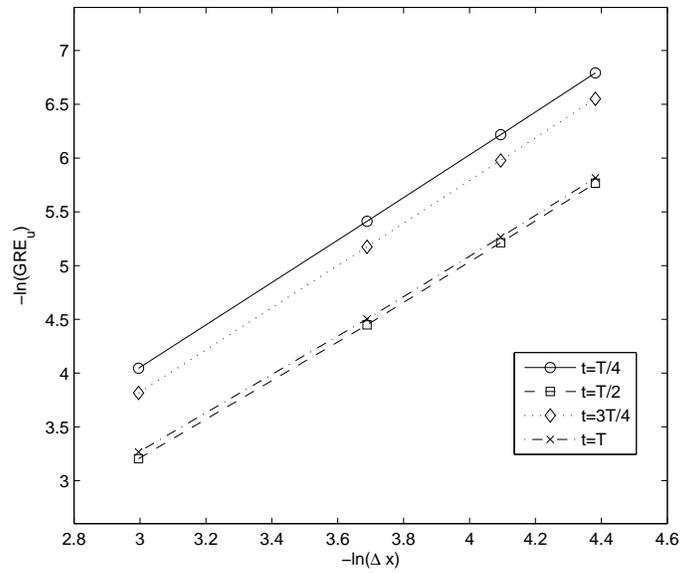}
\caption{The variation of $GRE_u$ with the lattice spacing for pulsatile flow at different times, $\eta=2.8285$.}
\label{edxw}
\end{figure}

\begin{figure}
\centering
\includegraphics[width=4.0in]{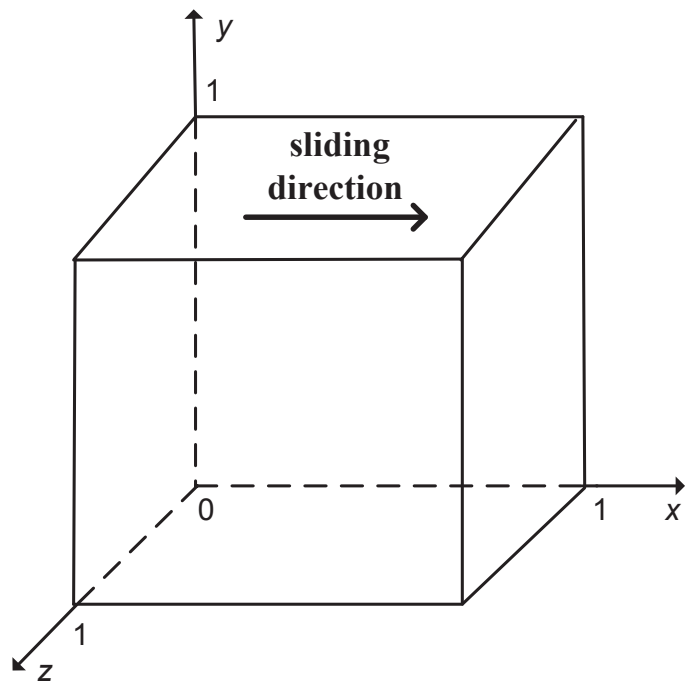}
\caption{The schematic of three-dimensional lid-driven cavity flow.}
\label{cavity-schematic}
\end{figure}

\begin{figure}
\centering (a) \\
\includegraphics[width=3.5in]{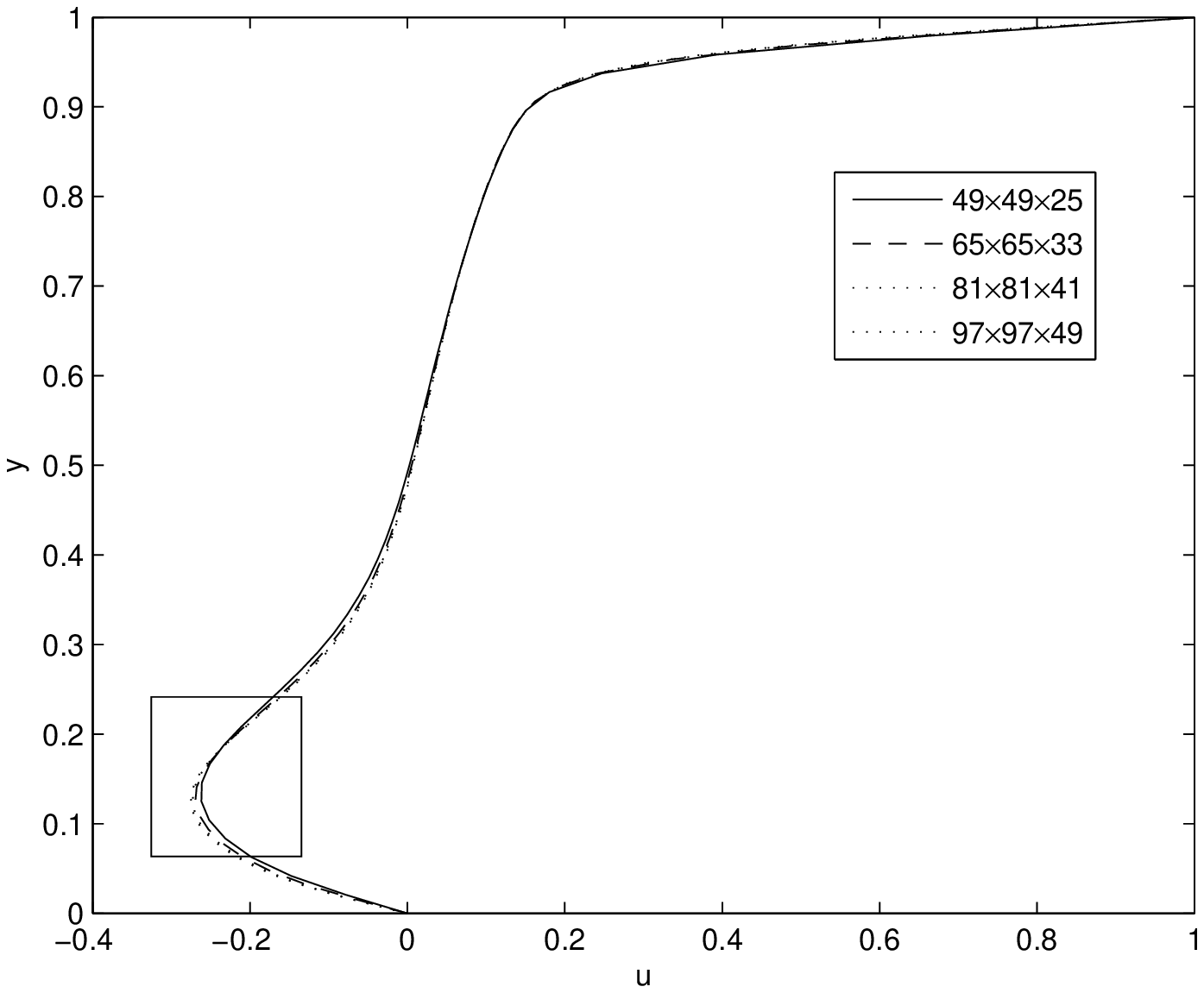}\\
 (b)\\
\includegraphics[width=3.5in]{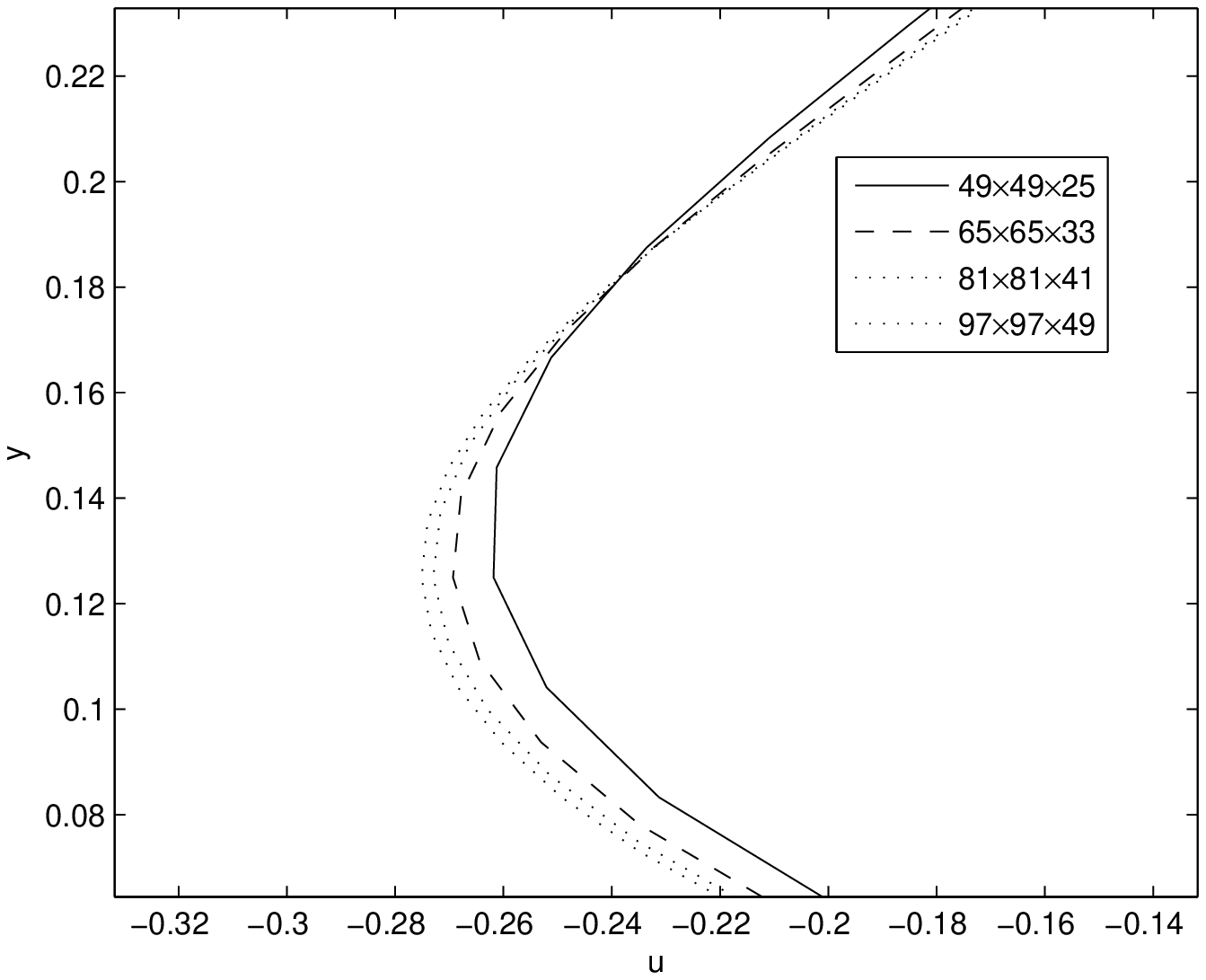}\\
\caption{The distribution of $u$ at four different grids in the vertical center line ($z=0.5$ and $x=0.5$) for cavity flow at Re=1000, (b) is the magnification of square area in (a).}
\label{cav_grid_independence}
\end{figure}

\begin{figure}
\centering (a) Re=100\\
\includegraphics[width=2.3in]{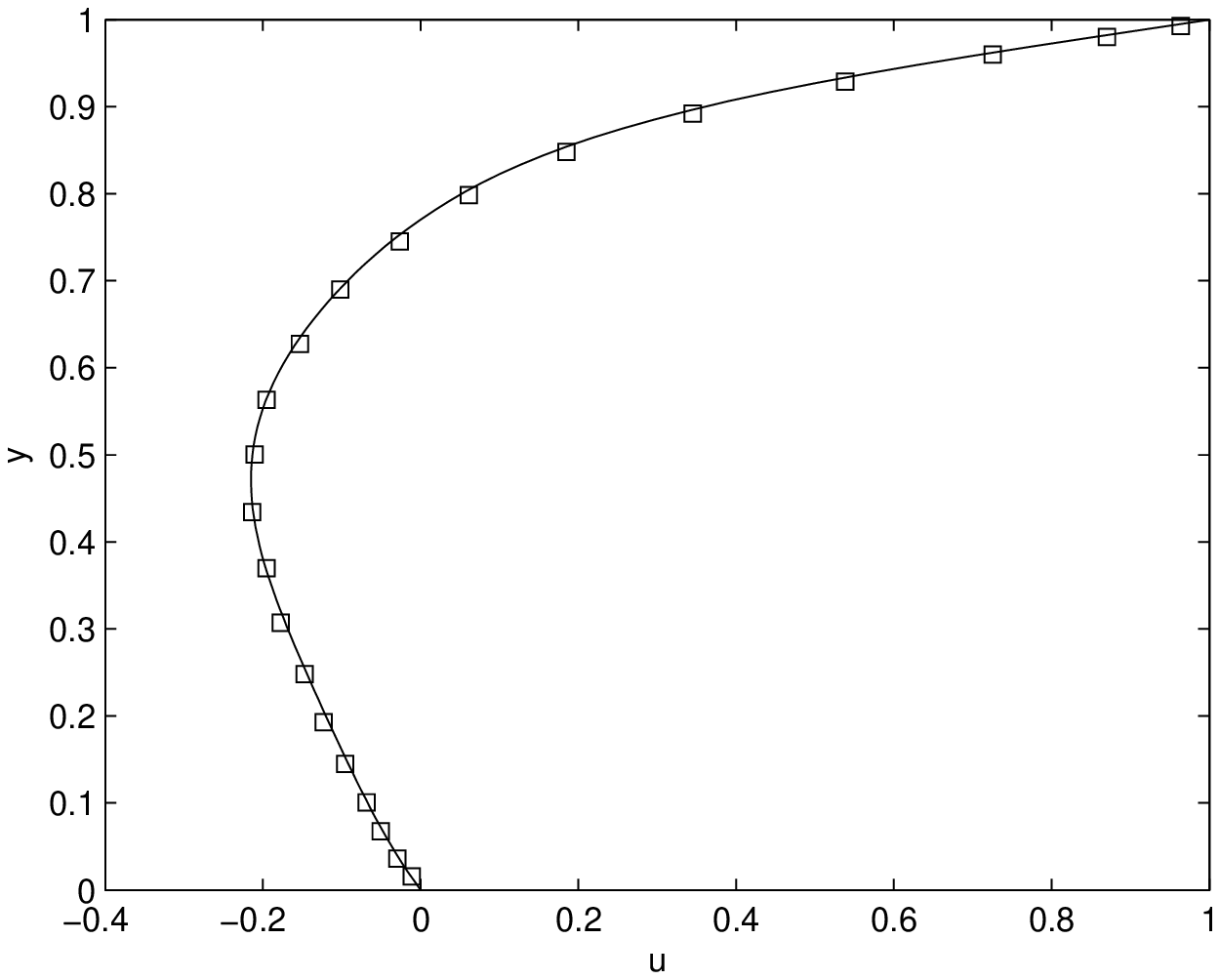}
\includegraphics[width=2.3in]{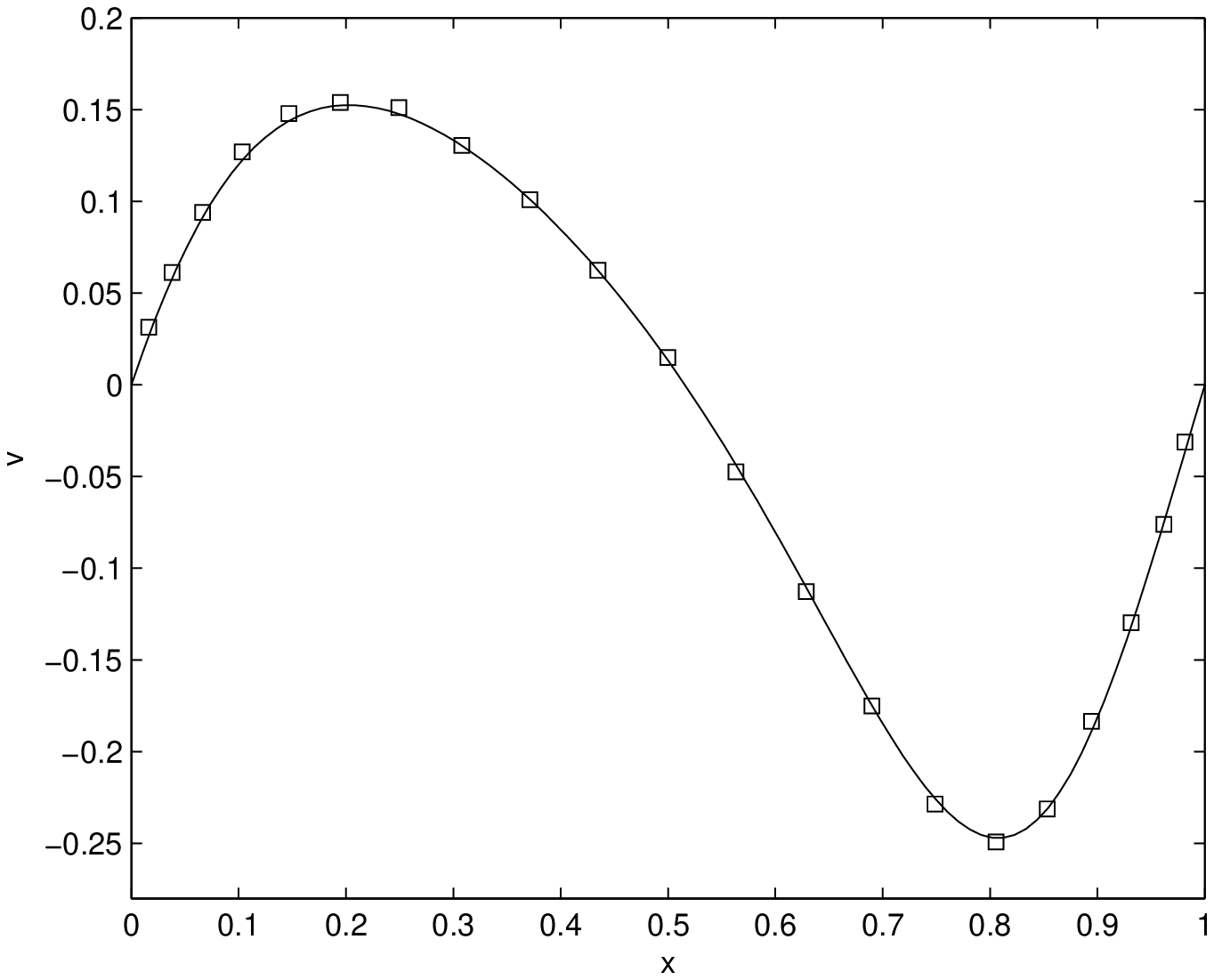}\\
(b) Re=400\\
\includegraphics[width=2.3in]{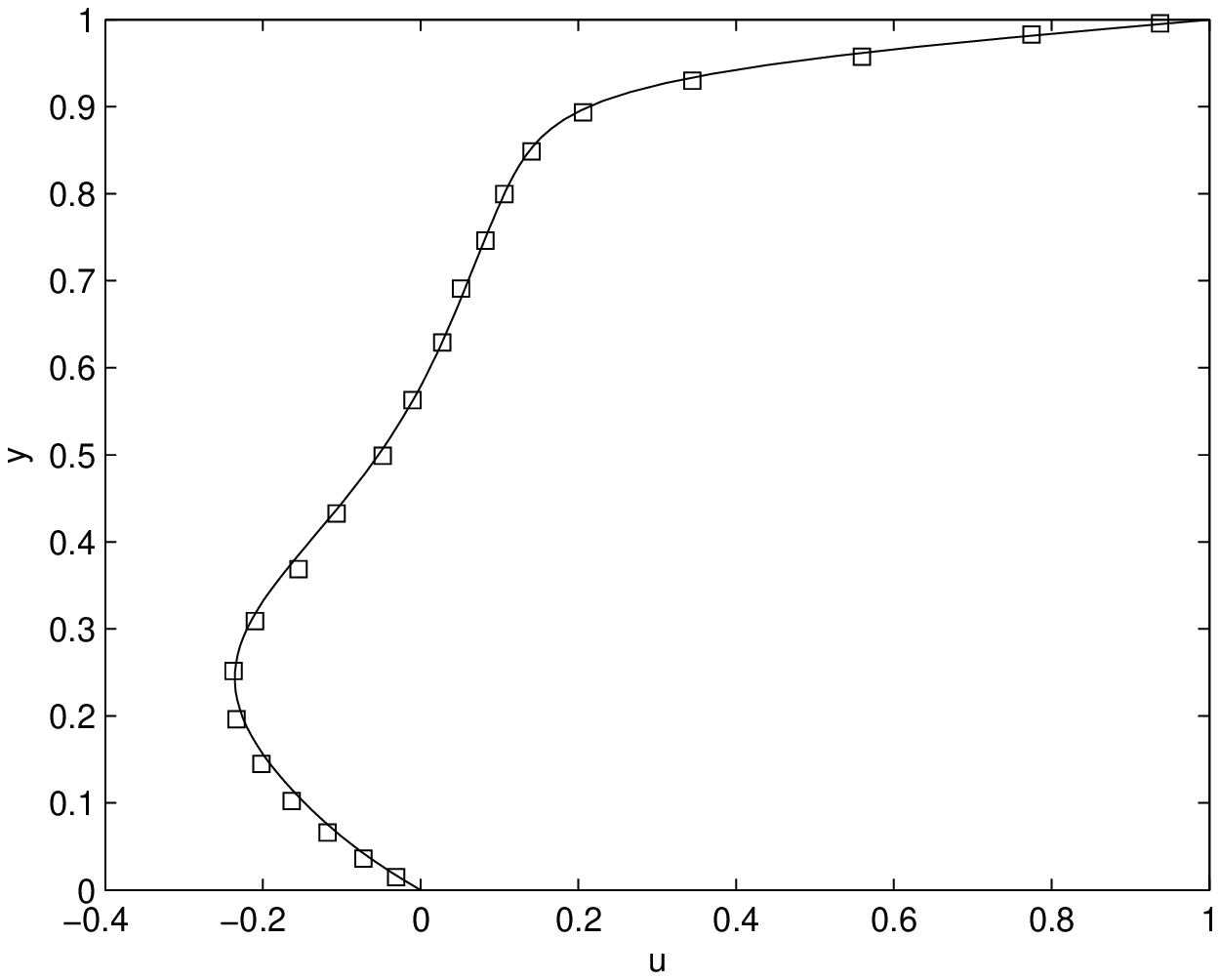}
\includegraphics[width=2.3in]{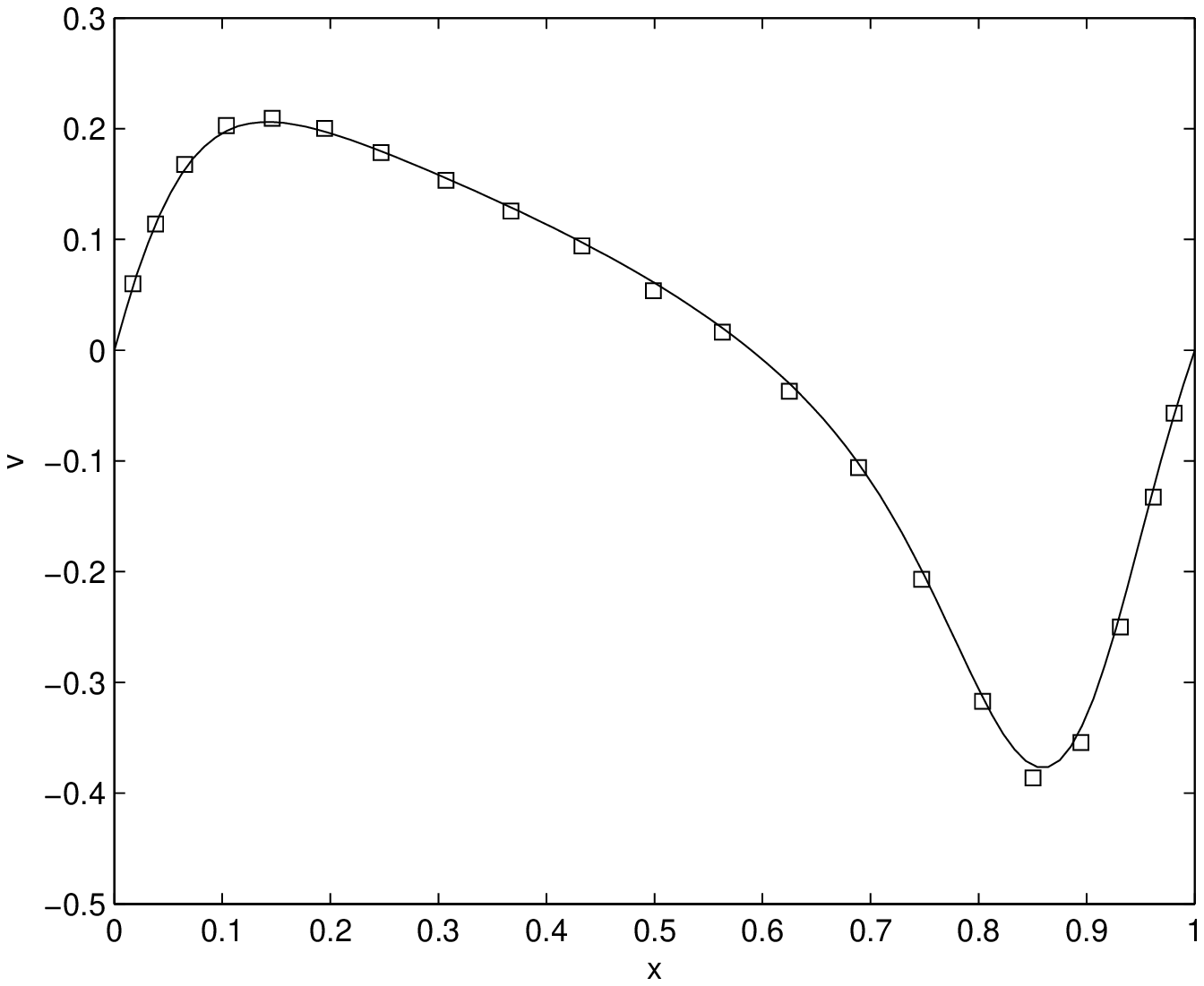}\\
(c) Re=1000\\
\includegraphics[width=2.3in]{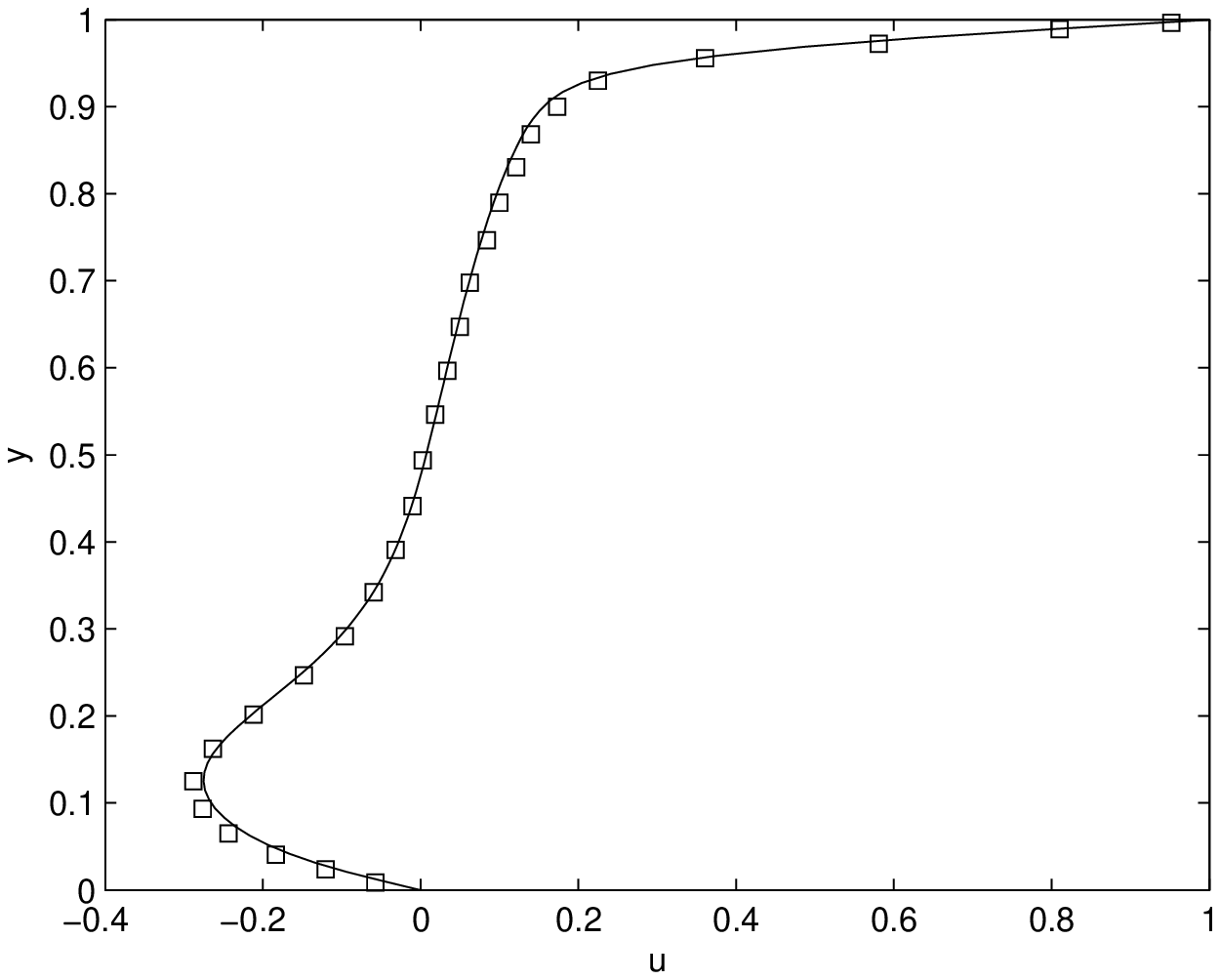}
\includegraphics[width=2.3in]{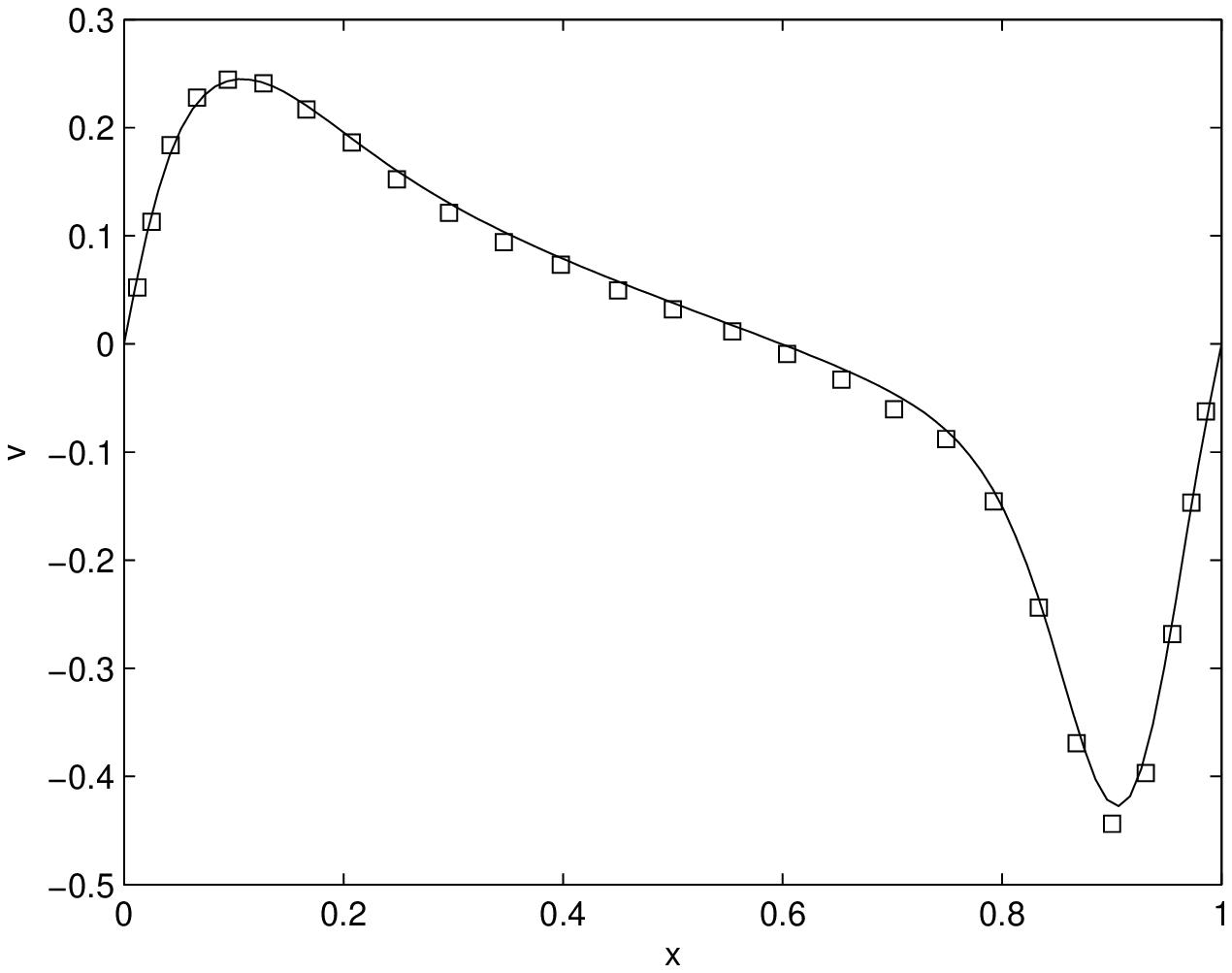}\\
\caption{The velocity distribution in the vertical and horizontal center lines at section $z=0.5$ for cavity flow at different Re, $\square$ : the results of Ku, -- : present simulation.}
\label{cavityplot}
\end{figure}

\begin{figure}
\centering
\includegraphics[width=3.5in]{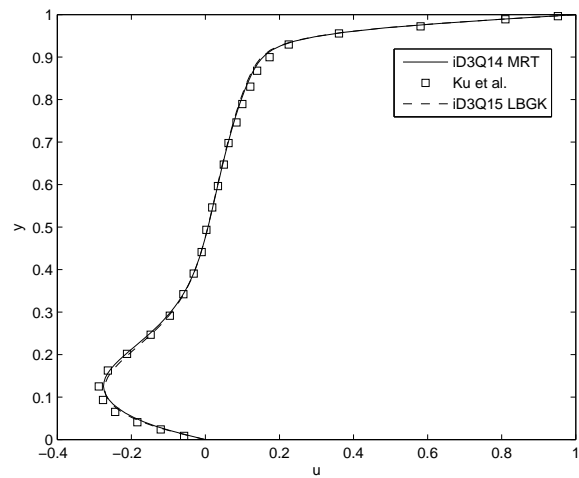}
\caption{The distribution of $u$ in vertical center line for cavity flow at Re=1000. The simulations are carried out with grid $97\times97\times49$.}
\label{imrt_vs_ibgk}
\end{figure}

\begin{figure}
\centering (a) iD3Q14 MRT model \\
\includegraphics[width=4.0in]{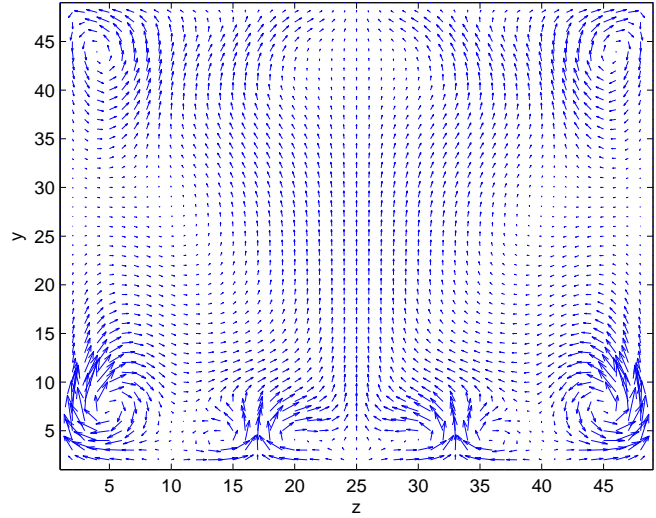}\\
(b) iD3Q15 LBGK model\\
\includegraphics[width=4.0in]{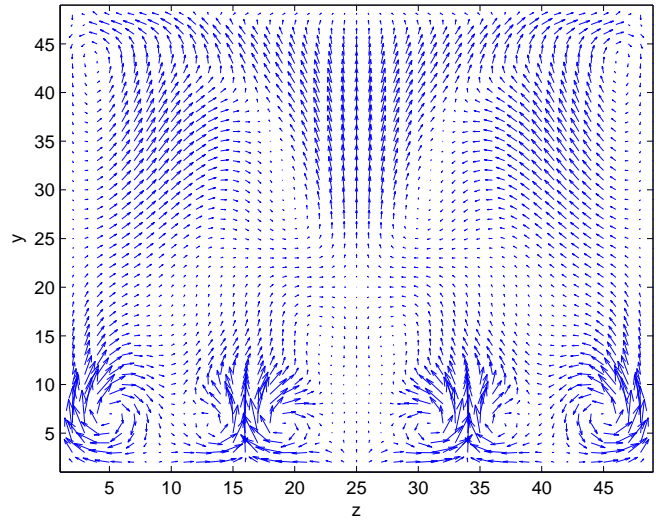}
\caption{The velocity vector plot of $yz$ plane at $x=0.5$ and $t=50000\delta t$ for cavity flow at Re=3200, the grid is $97\times97\times49$. The length of the arrows is three times the actual velocity magnitude. }
\label{yzplane_imrt_vs_ibgk}
\end{figure}

\begin{figure}
\centering iD3Q14 MRT model \\
\includegraphics[width=4.0in]{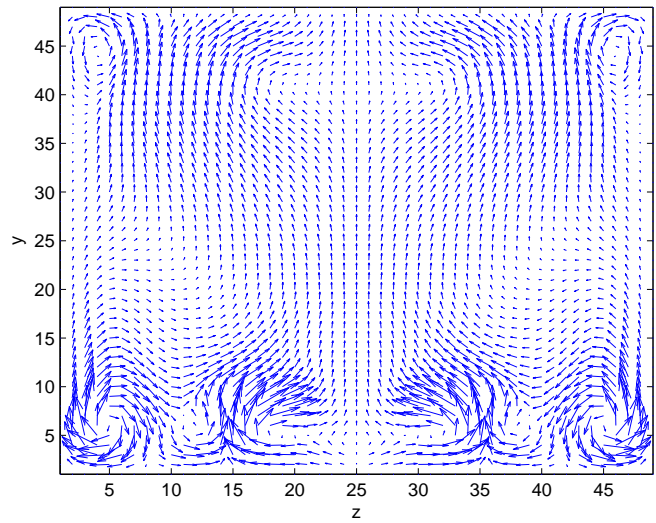}
\caption{The velocity vector on the $yz$ plane at $x=0.5$ and $t=25000\delta t$ for cavity flow at Re=3200, the grid is $49\times49\times25$. The length of the arrows is three times the actual velocity magnitude. The result is obtained from iD3Q14 MRT model, while the simulation by iD3Q15 LBGK model diverges.}
\label{cavity_imrt14_49}
\end{figure}

\end{document}